\documentclass{aa}  
\usepackage{graphicx}
\usepackage{mathtools}
\usepackage{amsmath}
\usepackage{fancyref}
\usepackage{longtable}
\usepackage{rotating}
\usepackage{pdflscape}
\usepackage{enumerate}
\usepackage[citecolor=blue, urlcolor=blue, linkcolor=blue, colorlinks=true]{hyperref}
\usepackage{txfonts}
\usepackage{natbib}
\begin{document}

   \title{Search for associations containing young stars (SACY)}

   \subtitle{VI. Is multiplicity universal? Stellar multiplicity in the range 3-1000\,au from adaptive-optics observations\thanks{Based on observations obtained using the instruments NACO at the VLT (077.C-0483, 081.C-0825, 088.C-0506, 089.C-0207) and the adaptive optics facilities at the Lick observatory (UCO 3m)}$^{,}$\thanks{Appendix~\ref{sec:appendix_a} is available in electronic form at \url{http://www.aanda.org}.  Table~\ref{tab:all_individ_detections} is only available at the CDS via anonymous ftp to \nolinkurl{http://cdsarc.u-strasbg.fr (ftp://123.45.678.9)} or via \url{http://cdsarc.u-strasbg.fr/viz-bin/qcat?J/A+A/XXX/XXX}}}

   \author{P. Elliott
          \inst{1, 2}\and
   	  N. Hu\'elamo\inst{3}
          \and
          H. Bouy\inst{3}
          \and
          A. Bayo
	  \inst{4}\and
	  C. H. F. Melo
	  \inst{1}\and
          C. A. O. Torres\inst{5}
          \and 
          M. F. Sterzik\inst{1}
          \and
          G. R. Quast\inst{5}     
          \and
          G. Chauvin\inst{6} 
          \and
          D. Barrado\inst{3} 
          }
   \institute{European Southern Observatory, Alonso de Cordova 3107, Vitacura Casilla 19001, Santiago 19, Chile
              \\
              \email{pe210@exeter.ac.uk}
         \and
            School of Physics, University of Exeter, Stocker Road, Exeter, EX4 4QL
	\and 
	   Centro de Astrobiolog\'{\i}a, ESAC Campus, Apdo. 78, E-28691 Villanueva de la Ca\~nada 			(Madrid), Spain
	\and
           Departamento de F\'isica y Astronom\'ia, Facultad de Ciencias, Universidad de Valpara\'iso, Av. Gran       Breta\~na 1111, 5030 Casilla, Valpara\'iso, Chile
	 \and
	    Laborat\'orio Nacional de Astrof\'isica/ MCT, Rua Estados Unidos 154, 37504-364 Itajub\'a (MG), Brazil
	\and
	Laboratoire d'Astrophysique, Observatoire de Grenoble, BP 53, 38041 Grenoble, Cedex 9, France
             }

   \date{Received 3 Feb 2015; accepted 28 May 2015}

  \abstract
   {Young loose nearby associations are unique samples of close ($<$150\,pc), young ($\approx$5-100\,Myr) pre-main sequence (PMS) stars. A significant number of members of these associations have been identified in the SACY collaboration.  We can use the proximity and youth of these members to investigate key ingredients in star formation processes, such as multiplicity. }
   {With the final goal to better understand multiplicity properties at different evolutionary stages of PMS stars, we present the statistics of identified multiple systems from 113 confirmed SACY members.  We derive multiplicity frequencies, mass-ratio,  and physical separation distributions in a consistent parameter space, and compare our results to other PMS populations and the field.}
   {We have obtained adaptive-optics assisted near-infrared observations with NACO (ESO/VLT) and IRCAL (Lick Observatory) for at least one epoch of all 113 SACY members. 
We have identified multiple systems using co-moving proper-motion analysis for targets with multi-epoch data,  and using contamination estimates in terms of mass-ratio and physical separation for targets with single-epoch data.  We have explored ranges in projected separation and mass-ratio of
$a$ [3--1000\,au], and $q$ [0.1--1], respectively.}
   {We have identified 31 multiple systems (28 binaries and 3 triples).  
We derive a multiplicity frequency (MF) of MF$_{3-1000\,\mathrm{au}}$=28.4$^{+4.7}_{-3.9}$\% and a triple frequency (TF) of TF$_{3-1000\,\mathrm{au}}$=2.8$^{+2.5}_{-0.8}$\% in the separation range of 3-1000\,au.  We do not find any evidence for an increase in the MF with primary mass. The estimated mass-ratio of our statistical sample (with power-law index $\gamma$=$-0.04\pm0.14$) is consistent with a flat distribution ($\gamma=0$). 
}
{Analysis from previous work using tight binaries indicated that the underlying multiple system distribution of the SACY dataset and the young star-forming region (SFR) Taurus are statistically similar, supporting the idea that these two populations formed in a similar way.  In this work, we show further similarities (but also hints of discrepancies) between the two populations: flat mass-ratio distributions and statistically similar MF and TF values. 
We also compared the SACY sample to the field (in the separation range of 19-100\,au), finding that the two distributions are indistinguishable, suggesting a similar formation mechanism.}


   \maketitle
%

\section{Introduction}

The multiplicity of a stellar population holds a huge amount of information about the birth, evolution and eventual fate of the population itself and its individual members.  Thus, comprehensively characterising these properties has implications across many fields such as star formation, N-body dynamics, and supernovae rates.  
Multiplicity surveys of solar-type stars \citep[e.g.][]{Duquennoy1991, Raghavan2010, Tokovinin2014} have made use of a combination of techniques (from small physical separations to large: spectroscopy, interferometry including sparse aperture masking, -- SAM --, speckle imaging, adaptive optics, -- AO --, assisted imaging, and classical direct imaging) to probe and characterise multiple systems with periods between $\sim$1-10$^{10}$ days (with a peak at $\approx$10$^5$\,day). 
As a result, they show that almost one out of every two solar-type star (0.9-1.5\,M$_\odot$) has one or more companion.  Considering the high-mass end of the initial mass function (IMF), the O-type stars, the percentage increases up to $\approx$100\% \citep{Sana2014}.  In other words, these massive stars form almost exclusively in multiple systems. 

Such surveys are volume-limited, making use of the high number of main sequence (MS) field stars that are nearby to produce robust statistics
 \citep[e.g., in the case of][90\% completeness, considering their spatial density, for $d<$ 70 pc]{Tokovinin2014}. However, the history of this population is not well constrained as it comprises stars born at different times and in a range of environments. As a result, they must have undergone quite different dynamical processing\citep{Parker2009}.  Therefore, the information regarding the multiplicity properties as a direct output from star formation has been smeared out during its somewhat unreproducible dynamical history.  To study a more pristine population one has to consider the observations of younger stars ($<$100\,Myr) in their nurseries, that usually provide lower number statistics. 
Such studies are usually performed in nearby star-forming regions (SFRs) like e.g. Taurus \citep{Leinert1993, Kraus2011, Daemgen2014}, Ophiuchus/L1688 \citep{Ratzka2005}, Chamaeleon I \citep{Lafreniere2008}, IC 348: \citep{Duchene1999}, or the ONC \citep{Reipurth2007}.  However, as \cite{King2012a} demonstrated, the amount of overlapping cluster-to-cluster parameter space limits the number of systems that one is able to statistically compare (typically $<$100).

Young associations have also been studied due to their proximity ($<$150\,pc) and youth ($\lesssim$100\,Myr).  There are a handful of studies focused on the stellar multiplicity in these populations \citep[see e.g.][]{Brandeker2003, Brandeker2006}.
The results of these works lead to varying conclusions about the nature of the stars in such associations.  As an example, \cite{Brandeker2003} found a very high frequency of stellar multiple systems in TW-Hydrae, and an absence of systems (with separations $>$20\,au) in $\eta$ Chamaeleontis, with an extremely small likelihood ($<10^{-4}$) that the two populations share the same parental multiplicity distribution. It is still unclear whether this is due to dependencies on purely initial density \citep{Moraux2007} or additional environmental effects \citep{Becker2013}.

There have also been some studies focused on the substellar multiplicity in these associations \citep{Masciadri2005, Lafreniere2007, Nielsen2010, Evans2012, Chauvin2010, Biller2013}.  In addition to such surveys, many individual objects have been imaged with the aim to disentangle the planetary / brown dwarf nature of their companions and further understand their circumstellar discs \citep{Chauvin2004, Huelamo2011}.  

However, a detailed statistical analysis of the abundance and characteristics of multiple systems in young associations compared to those identified in SFRs is still missing.  Such an analysis was very complex until recently, posing a compromise between the publicly available data and the ability to probe the same parameter space.  
Motivated by this, in 2006 we started an homogeneous observational campaign to derive the multiplicity of young loose associations. 

We aim to characterise the multiplicity properties of these associations across a large parameter space ($a\sim$ 0.1-10$^4$\,au) by combining a range of observational techniques.  We have already derived the spectroscopic multiple system fraction (the tightest multiple systems, semi-major axis ($a$) $\lesssim$1\,au) of our sample \citep{Elliott2014}. We found that the derived multiplicity fraction is compatible with that of the field and nearby SFRs, as one would expect due to the lack of dynamical processing for systems with such high binding energy, that have statistically similar primordial abundances.  We are now extending our analysis to wider binaries, using AO imaging.

AO imaging allows us to identify and characterise multiple systems across a huge parameter space for targets that belong to nearby associations.  As an example, for an {\it average} member of our sample ($d$=70\,pc, age=30\,Myr, $M_1$=1\,M$_{\odot}$),  an instrument such as NAOS-CONICA at the VLT (point spread function, --PSF--, full-width half maximum, --FWHM--, of 0.07\,$\arcsec$ at 2.2\,$\mu$m, with a strehl ratio of 1), is sensitive to companions at separations  [$a$\,$\sim$3-1000\,au] in the [$M_2\sim$ 0.01-1\,M$_{\odot}$] mass range.
 As mentioned previously, the peak of field stars period distribution is at 10$^5$\,day which, considering a secondary of 0.1\,M$_{\odot}$, corresponds to a semi-major axis of $\sim$40\,au assuming a circular orbit.  Therefore, our observing campaign should yield a high number of multiple systems when a large enough sample is observed. In addition, the observations must have sufficient temporal coverage to discard that a visual pair is just a chance alignment through a proper motion analysis (as orbital motion is negligible in the majority of cases). 

The manuscript is organised as follows:  Section~\ref{sec:sample_selec} describes the target selection.  Section~\ref{sec:Observations and Data} details the observations, data reduction, and the sensitivity limits, as well as the literature analysis.  Section~\ref{sec:Identifying multiple systems} outlines the methods used to identify bound multiple systems and the uncertainty estimations.  Section~\ref{section:Results} describes how we derived the multiplicity properties of our sample.  Section~\ref{sec:Discussion} presents the statistical analysis of the built distributions and comparisons to other populations. Section~\ref{section:conclusions} states the main conclusions of this work and prospects.

\section{Sample definition}

The search for associations containing young stars (SACY) dataset contains $\approx$\,2000 stars located all over the sky.  The stars are defined as members of distinct moving groups by the convergence method \citep[see census of][]{Torres2006}.  The kinematic properties of the stars are derived from the parallax, the proper motion (PM) and the radial velocity (RV) values.

The starting point for our sample selection was based on the census by \citet{Torres2006}.
For the southern targets two additional criteria were followed: (i) we only consider stars with spectral types later than G5, in order to increase the contrast between the primary and the possible companion candidate, and (ii) the right ascension (RA) should be obtainable from Paranal observatory when our service observations were conducted (periods 77, 81, 88, and 89). The northern targets also followed criterion (i) and were also observed according to their visibility during the two runs at the Lick observatory in 2008 and 2009. 

Later on, the membership of several targets was revised based on new data (e.g. new RV 
values derived from high-resolution spectra, \cite{Elliott2014}.  This revision allowed us to eliminate contamination within the sample from interlopers.  From the 201 SACY targets observed originally with AO, 113 are now confirmed SACY members, while 88 have been rejected. 

The bulk of this work is focused on the 113 confirmed SACY members, and the specific details of the sample are provided below (summarized in Table~\ref{tab:all_sacy_observed_targets}). However, for completeness with our original observations, we also provide details on the 88 non-SACY targets, including their observational parameters and a summary of the identified multiple systems, in Tables~\ref{tab:all_individ_detections} and ~\ref{tab:binaries_2}, respectively.



%
\subsection{Filtered coherent sample for this work}
\label{sec:sample_selec}

Table~\ref{tab:all_sacy_observed_targets} provides the census of the 113 confirmed SACY members (once the additional data from  \citealt{Elliott2014} have been included to identify interlopers) studied in this work.  We have included two K magnitude values: the one from the 2MASS survey, K (2MASS), and our estimated K (M$_1$).
From our observations (with a narrow field of view that does not allow for a robust flux calibration), we can only obtain relative photometry and therefore we use 2MASS magnitudes for calibration.  We denoted our estimated magnitude,  K (M$_1$), and it only differs from the original 2MASS value in the case of a detection of a new companion (previously unresolved) within the 2MASS PSF.  The typical PSF width for 2MASS is $\approx$2\,\arcsec and, in some cases, multiple components can be found within this angular distance.  To estimate the photometry of the primary and components 
we weighted the 2MASS value according to the observed flux ratio.  In the next sections, the primary masses have been  calculated using this value, and the age value included in Table~\ref{tab:summary_associations} with the evolutionary tracks of \cite{Baraffe1998}.

Finally, we note that all the targets in this sample have PM values from the UCAC4 catalogue \citep{Zacharias2012}.  Precise PMs will allow us to determine whether identified companions are co-moving with the primary source.

{
\onecolumn
\LTcapwidth=\textwidth
\begin{longtable}{p{5cm} p{1.5cm} p{1.5cm} p{0.7cm} p{0.7cm} p{0.9cm} p{0.7cm} p{0.7cm}}
\caption{All SACY targets observed and analysis in this work. The distance values are kinematic, derived from the convergence method \citep{Torres2006}.}
\label{tab:all_sacy_observed_targets}\\
\hline\hline\\[-1ex]
  \multicolumn{1}{l}{ID} &
  \multicolumn{1}{l}{RA} &
  \multicolumn{1}{l}{DEC} &
  \multicolumn{1}{l}{K (2MASS)} &
  \multicolumn{1}{l}{K (M$_1$)} &
  \multicolumn{1}{l}{Ass.} &
  \multicolumn{1}{l}{Distance} &
  \multicolumn{1}{l}{M$_1$} \\
  \multicolumn{1}{l}{} &
  \multicolumn{1}{l}{hh:mm:ss.s} &
  \multicolumn{1}{l}{dd:mm:ss} &
  \multicolumn{1}{l}{(mag.)} &
  \multicolumn{1}{l}{(mag.)} &
  \multicolumn{1}{l}{} &
  \multicolumn{1}{l}{(pc)} &
  \multicolumn{1}{l}{(M$_\odot$)} \\[-0.1ex]
\hline\\[-1ex]
\endfirsthead
\hline\hline\\[-1ex]
  \multicolumn{1}{l}{ID} &
  \multicolumn{1}{l}{RA} &
  \multicolumn{1}{l}{DEC} &
  \multicolumn{1}{l}{K (2MASS)} &
  \multicolumn{1}{l}{K (M$_1$)} &
  \multicolumn{1}{l}{Ass.} &
  \multicolumn{1}{l}{Distance} &
  \multicolumn{1}{l}{M$_1$} \\
  \multicolumn{1}{l}{} &
  \multicolumn{1}{l}{hh mm ss} &
  \multicolumn{1}{l}{dd mm ss} &
  \multicolumn{1}{l}{(mag.)} &
  \multicolumn{1}{l}{(mag.)} &
  \multicolumn{1}{l}{} &
  \multicolumn{1}{l}{(pc)} &
  \multicolumn{1}{l}{(M$_\odot$)} \\[-0.1ex]
\hline\\[-1ex]
\endhead
  CD-45 14955B            & 23:11:53.6 & -45:08:00 & 8.85 & 8.85 & ABD   & 49.3 & 0.56\\
  GJ 9809                 & 23:06:04.8 & +63:55:34 & 6.98 & 6.98 & ABD   & 24.5 & 0.61\\
  HD 1405                 & 00:18:20.9 & +30:57:22 & 6.51 & 6.51 & ABD   & 27.3 & 0.78\\
  HD 160934               & 17:38:39.6 & +61:14:16 & 6.81 & 6.81 & ABD   & 33.1 & 0.8\\
  HD 201919               & 21:13:05.2 & -17:29:12 & 7.58 & 7.58 & ABD   & 39.5 & 0.72\\
  HD 222575               & 23:41:54.2 & -35:58:39 & 7.62 & 7.62 & ABD   & 63.6 & 0.96\\
  HD 24681                & 03:55:20.4 & -01:43:45 & 7.25 & 7.25 & ABD   & 53.3 & 0.96\\
  LP 745-70               & 16:33:41.6 & -09:33:11 & 7.55 & 7.55 & ABD   & 30.3 & 0.59\\
  TYC 486-4943-1          & 19:33:03.7 & +03:45:39 & 8.66 & 8.66 & ABD   & 72.0 & 0.77\\
  V* PX Vir               & 13:03:49.6 & -05:09:42 & 5.51 & 5.51 & ABD   & 21.7 & 0.9\\
  HD 152555               & 16:54:08.1 & -04:20:24 & 6.36 & 6.36 & ABD   & 46.8 & 1.13\\
  GJ 885 A                & 23:00:27.9 & -26:18:42 & 6.27 & 6.75 & ABD   & 32.7 & 0.81\\
  Wolf 1225               & 22:23:29.1 & +32:27:34 & 6.05 & 6.77 & ABD   & 15.1 & 0.48\\
  BD-03 4778              & 20:04:49.3 & -02:39:20 & 7.92 & 8.06 & ABD   & 71.5 & 0.91\\
  GJ 4231                 & 21:52:10.4 & +05:37:35 & 7.38 & 8.2 & ABD   & 30.6 & 0.5\\
  TYC 91-82-1             & 04:37:51.4 & +05:03:08 & 8.65 & 8.86 & ABD   & 87.6 & 0.82\\
  CD-52 9381              & 20:07:23.7 & -51:47:27 & 7.39 & 7.39 & ARG   & 29.4 & 0.59\\
  2MASS J05200029+0613036 & 05:20:00.2 & +06:13:03 & 8.57 & 8.57 & BPC   & 67.8 & 0.67\\
  BD-13 6424              & 23:32:30.8 & -12:15:51 & 6.57 & 6.57 & BPC   & 27.9 & 0.7\\
  CD-31 16041             & 18:50:44.4 & -31:47:47 & 7.46 & 7.46 & BPC   & 53.2 & 0.87\\
  CD-54 7336              & 17:29:55.0 & -54:15:48 & 7.36 & 7.36 & BPC   & 68.4 & 1.06\\
  CPD-72 2713             & 22:42:48.9 & -71:42:21 & 6.89 & 6.89 & BPC   & 37.3 & 0.81\\
  GSC 07396-00759         & 18:14:22.0 & -32:46:10 & 8.54 & 8.54 & BPC   & 95.2 & 0.93\\
  HD 161460               & 17:48:33.7 & -53:06:43 & 6.78 & 6.78 & BPC   & 71.0 & 1.2\\
  Smethells 20            & 18:46:52.5 & -62:10:36 & 7.85 & 7.85 & BPC   & 52.4 & 0.73\\
  TYC 6872-1011-1         & 18:58:04.1 & -29:53:04 & 8.02 & 8.02 & BPC   & 82.6 & 0.99\\
  V* TX PsA               & 22:45:00.0 & -33:15:25 & 7.79 & 7.79 & BPC   & 20.1 & 0.19\\
  V* V4046 Sgr            & 18:14:10.4 & -32:47:34 & 7.25 & 7.25 & BPC   & 76.9 & 1.15\\
  V* WW PsA               & 22:44:57.9 & -33:15:01 & 6.93 & 6.93 & BPC   & 20.1 & 0.35\\
  GJ 3322                 & 05:01:58.7 & +09:58:59 & 6.37 & 6.8 & BPC   & 37.8 & 0.85\\
  CD-26 13904             & 19:11:44.6 & -26:04:08 & 7.37 & 7.77 & BPC   & 78.9 & 1.03\\
  CD-27 11535             & 17:15:03.6 & -27:49:39 & 7.38 & 8.06 & BPC   & 87.3 & 1.01\\
  GSC 08350-01924         & 17:29:20.6 & -50:14:53 & 7.99 & 8.65 & BPC   & 66.3 & 0.63\\
  2MASS J05203182+0616115 & 05:20:31.8 & +06:16:11 & 8.57 & 8.7 & BPC   & 71.0 & 0.66\\
  2MASS J08110934-5555563 & 08:11:09.3 & -55:55:56 & 9.4 & 9.4 & CAR   & 129.0 & 0.84\\
  2MASS J08521921-6004443 & 08:52:19.2 & -60:04:44 & 9.37 & 9.37 & CAR   & 163.9 & 1.04\\
  2MASS J08563149-5700406\tablefoottext{a} & 08:56:31.4 & -57:00:40 & 9.76 & 9.76 & CAR   & 191.6 & 0.94\\
  BD-03 5579              & 23:09:37.1 & -02:25:55 & 7.82 & 7.82 & CAR   & 63.0 & 0.85\\
  CD-44 1533              & 04:22:45.6 & -44:32:51 & 8.58 & 8.58 & CAR   & 103.7 & 0.91\\
  CD-49 4008              & 08:57:52.1 & -49:41:50 & 8.64 & 8.64 & CAR   & 102.9 & 0.89\\
  CD-53 2515              & 08:51:56.4 & -53:55:56 & 8.75 & 8.75 & CAR   & 137.1 & 1.13\\
  CD-54 2499              & 08:59:28.7 & -54:46:49 & 8.4 & 8.4 & CAR   & 109.5 & 1.08\\
  CD-55 2543              & 09:09:29.3 & -55:38:27 & 8.4 & 8.4 & CAR   & 115.2 & 1.12\\
  CD-61 2010              & 08:42:00.4 & -62:18:26 & 8.83 & 8.83 & CAR   & 121.4 & 0.93\\
  CD-69 783               & 10:41:23.0 & -69:40:43 & 8.38 & 8.38 & CAR   & 86.5 & 0.87\\
  HD 107722               & 12:23:29.0 & -77:40:51 & 7.14 & 7.14 & CAR   & 59.0 & 1.05\\
  HD 309751               & 09:31:44.7 & -65:14:52 & 8.36 & 8.36 & CAR   & 141.6 & 1.2\\
  HD 8813                 & 01:23:25.8 & -76:36:42 & 6.79 & 6.79 & CAR   & 46.5 & 0.92\\
  TYC 8174-1586-1         & 09:11:15.8 & -50:14:14 & 9.5 & 9.5 & CAR   & 117.2 & 0.79\\
  TYC 8557-1251-1         & 07:55:31.6 & -54:36:50 & 9.19 & 9.19 & CAR   & 123.0 & 0.86\\
  TYC 8582-3040-1\tablefoottext{a}         & 08:57:45.6 & -54:08:36 & 9.35 & 9.35 & CAR   & 143.2 & 0.89\\
  TYC 8962-1747-1         & 11:08:07.9 & -63:41:47 & 8.29 & 8.29 & CAR   & 88.2 & 0.9\\
  TYC 9217-417-1          & 09:59:57.6 & -72:21:47 & 8.69 & 8.69 & CAR   & 83.6 & 0.8\\
  TYC 9486-927-1          & 21:25:27.4 & -81:38:27 & 7.34 & 7.34 & CAR   & 36.3 & 0.71\\
  HD 22213                & 03:34:16.3 & -12:04:07 & 6.79 & 6.97 & CAR   & 52.4 & 0.94\\
  BD-07 2388              & 08:13:50.9 & -07:38:24 & 6.92 & 7.4 & CAR   & 42.6 & 0.77\\
  2MASS J09131689-5529032 & 09:13:16.8 & -55:29:03 & 8.36 & 8.61 & CAR   & 131.7 & 1.15\\
  CD-57 1709              & 07:21:23.7 & -57:20:37 & 8.7 & 8.7 & CAR   & 100.5 & 0.87\\
  CPD-62 1293             & 09:43:08.8 & -63:13:04 & 8.6 & 8.82 & CAR   & 69.4 & 0.7\\
  2MASS J08371096-5518105 & 08:37:10.9 & -55:18:10 & 9.38 & 9.38 & CAR   & 180.8 & 1.11\\
  2MASS J03573723-0416159 & 03:57:37.2 & -04:16:15 & 8.75 & 8.75 & COL   & 106.5 & 0.89\\
  2MASS J09322609-5237396 & 09:32:26.0 & -52:37:39 & 8.84 & 8.84 & COL   & 101.9 & 0.85\\
  BD+08 742               & 04:42:32.0 & +09:06:00 & 9.12 & 9.12 & COL   & 105.6 & 0.81\\
  BD-11 648\tablefoottext{a}               & 03:21:49.6 & -10:52:17 & 9.26 & 9.26 & COL   & 128.1 & 0.86\\
  CD-36 1785              & 04:34:50.7 & -35:47:21 & 8.59 & 8.59 & COL   & 80.1 & 0.8\\
  HD 26980                & 04:14:22.5 & -38:19:01 & 7.62 & 7.62 & COL   & 80.7 & 1.12\\
  HD 27679                & 04:21:10.3 & -24:32:20 & 7.81 & 7.81 & COL   & 78.0 & 0.94\\
  HD 31242                & 04:51:53.5 & -46:47:13 & 8.16 & 8.16 & COL   & 68.1 & 0.82\\
  TYC 9178-284-1          & 06:55:25.1 & -68:06:21 & 8.94 & 8.94 & COL   & 108.3 & 0.86\\
  V* V479 Car             & 09:23:34.9 & -61:11:35 & 7.96 & 7.96 & COL   & 87.7 & 1.07\\
  2MASS J02303239-4342232 & 02:30:32.4 & -43:42:23 & 7.23 & 7.57 & COL   & 51.3 & 0.81\\
  V* V1221 Tau            & 03:28:14.9 & +04:09:47 & 7.44 & 8.01 & COL   & 81.5 & 0.92\\
  2MASS J08240598-6334024 & 08:24:05.9 & -63:34:02 & 8.13 & 8.18 & COL   & 118.6 & 1.2\\
  HD 272836               & 04:53:05.1 & -48:44:38 & 8.24 & 8.34 & COL   & 78.6 & 0.84\\
  BD-16 351               & 02:01:35.6 & -16:10:00 & 7.96 & 8.43 & COL   & 80.9 & 0.83\\
  2MASS J04272050-4420393 & 04:27:20.4 & -44:20:39 & 8.56 & 8.7 & COL   & 86.2 & 0.81\\
  CD-43 1395              & 04:21:48.6 & -43:17:32 & 8.42 & 9.16 & COL   & 141.0 & 0.93\\
  2MASS J11594226-7601260 & 11:59:42.2 & -76:01:26 & 8.3 & 8.3 & ECH   & 101.0 & 0.98\\
  2MASS J12194369-7403572 & 12:19:43.6 & -74:03:57 & 8.86 & 8.86 & ECH   & 108.9 & 0.78\\
  2MASS J12392124-7502391 & 12:39:21.2 & -75:02:39 & 7.78 & 7.78 & ECH   & 108.6 & 1.2\\
  CD-69 1055              & 12:58:25.5 & -70:28:49 & 7.55 & 7.55 & ECH   & 106.7 & 1.2\\
  HD 104467               & 12:01:39.1 & -78:59:16 & 6.85 & 6.85 & ECH   & 106.5 & 1.2\\
  V* DZ Cha               & 11:49:31.8 & -78:51:01 & 8.49 & 8.49 & ECH   & 102.7 & 0.91\\
  V* MP Mus               & 13:22:07.5 & -69:38:12 & 7.29 & 7.29 & ECH   & 106.1 & 1.2\\
  HD 105923               & 12:11:38.1 & -71:10:36 & 7.18 & 7.26 & ECH   & 117.1 & 1.2\\
  TYC 9245-535-1          & 12:56:08.3 & -69:26:53 & 7.99 & 8.34 & ECH   & 119.3 & 1.14\\
  2MASS J12202177-7407393 & 12:20:21.7 & -74:07:39 & 8.37 & 8.4 & ECH   & 114.8 & 1.07\\
  2MASS J05581182-3500496 & 05:58:11.8 & -35:00:49 & 9.38 & 9.38 & OCT   & 105.8 & 0.77\\
  2MASS J06400573-3033089 & 06:40:05.7 & -30:33:08 & 8.7 & 8.7 & OCT   & 162.4 & 1.2\\
  CD-58 860               & 04:11:55.6 & -58:01:47 & 8.36 & 8.36 & OCT   & 87.1 & 0.88\\
  CD-66 395               & 06:25:12.3 & -66:29:10 & 9.0 & 9.0 & OCT   & 126.1 & 0.91\\
  HD 271037\tablefoottext{a}               & 05:06:50.5 & -72:21:11 & 8.67 & 8.67 & OCT   & 145.8 & 1.19\\
  BD-18 4452A             & 17:13:11.6 & -18:34:25 & 6.48 & 6.48 & OCT   & 16.7 & 0.51\\
  CD-30 3394A             & 06:40:04.9 & -30:33:03 & 8.59 & 8.59 & OCT   & 162.4 & 1.2\\
  CD-47 1999              & 05:43:32.1 & -47:41:10 & 8.64 & 8.68 & OCT   & 166.7 & 1.2\\
  BD-20 1111              & 05:32:29.3 & -20:43:33 & 8.7 & 8.7 & OCT   & 129.7 & 1.1\\
  HD 274576               & 05:28:51.3 & -46:28:19 & 8.81 & 8.92 & OCT   & 117.1 & 0.89\\
  2MASS J04302731-4248466 & 04:30:27.3 & -42:48:46 & 8.73 & 9.42 & OCT   & 122.8 & 0.82\\
  2MASS J06033540-4911256 & 06:03:35.4 & -49:11:25 & 9.08 & 9.81 & OCT   & 173.8 & 0.89\\
  2MASS J05490656-2733556 & 05:49:06.5 & -27:33:55 & 8.25 & 8.25 & THA   & 74.3 & 0.86\\
  ASAS J051536-0930.8     & 05:15:36.4 & -09:30:51 & 8.08 & 8.08 & THA   & 77.5 & 0.91\\
  BD-12 943               & 04:36:47.1 & -12:09:20 & 7.76 & 7.76 & THA   & 62.3 & 0.88\\
  BD-19 1062              & 04:59:32.0 & -19:17:41 & 8.07 & 8.07 & THA   & 64.1 & 0.83\\
  CD-35 1167              & 03:19:08.6 & -35:07:00 & 7.72 & 7.72 & THA   & 44.6 & 0.72\\
  CD-46 1064              & 03:30:49.0 & -45:55:57 & 7.1 & 7.1 & THA   & 42.2 & 0.84\\
  CD-53 544               & 02:41:46.8 & -52:59:52 & 6.76 & 6.76 & THA   & 41.4 & 0.9\\
  CD-78 24                & 00:42:20.3 & -77:47:39 & 7.53 & 7.53 & THA   & 50.1 & 0.83\\
  TYC 8098-414-1          & 05:33:25.5 & -51:17:13 & 8.16 & 8.16 & THA   & 51.6 & 0.68\\
  TYC 9344-293-1          & 23:26:10.7 & -73:23:49 & 7.94 & 7.94 & THA   & 44.2 & 0.65\\
  HD 22705                & 03:36:53.4 & -49:57:28 & 6.14 & 6.18 & THA   & 43.3 & 1.04\\
  CD-44 1173              & 03:31:55.6 & -43:59:13 & 7.47 & 7.58 & THA   & 44.0 & 0.75\\
  TYC 8083-455-1          & 04:48:00.6 & -50:41:25 & 7.92 & 8.25 & THA   & 54.8 & 0.7\\
  2MASS J05182904-3001321 & 05:18:29.0 & -30:01:32 & 8.3 & 8.41 & THA   & 66.2 & 0.76\\
\hline\end{longtable}
\tablefoot{\tablefoottext{a}{Targets excluded from statistical analysis due to the quality of their sensitivity curves, see Figure~\ref{fig:ang_sep}.}}

}
\twocolumn


\section{Observations and data reduction}
\label{sec:Observations and Data}

\subsection{Southern Targets: NACO data}

We had four sets of observations in service mode dedicated to collecting data on SACY sources with NAOS-CONICA, an adaptive optics system and near-IR camera at the very large telescope (VLT). The first epoch of data was taken between March--August 2006 (077.C-0483), and 76 targets were observed (seeing $\approx$0.9$\arcsec$).  The 40 objects (out of the 76) identified to have potential companions were re-observed (seeing $\approx$0.7$\arcsec$) between March--June 2008 (081.C-0825). We had two additional programs to observe other members of the SACY associations in 2011 and 2012 (088.C-0506, and 089.C-0207), and
we observed 60 and 25 targets (seeing $\approx$1.0$\arcsec$), respectively (see Table~\ref{tab:all_targets} for details on individual sources).

{\begin{table*}
\centering
\caption{Summary of the properties and frequency of visual multiple systems identified for the SACY associations studied in this work.  The minimum and maximum ages represent the dispersion found in the literature.
}
\begin{tabular}{p{2.0cm} p{0.8cm} p{1.3cm} p{1.3cm} p{1.3cm} p{1.3cm} p{1.3cm} p{1.3cm}}
\hline\hline\\
  \multicolumn{1}{l}{Name} &
  \multicolumn{1}{l}{ID} &
  \multicolumn{1}{l}{Dist.} &
  \multicolumn{1}{l}{Age\tablefoottext{a}} &
  \multicolumn{1}{l}{Max. / Min.} &
  \multicolumn{1}{l}{N. of observed\tablefoottext{b}} &
  \multicolumn{1}{l}{N. of VBs\tablefoottext{c}} &
   \multicolumn{1}{l}{MF$_{3-1000\,\mathrm{au}}$\tablefoottext{d}}\\
  \multicolumn{1}{l}{} &
  \multicolumn{1}{l}{} &
  \multicolumn{1}{l}{(pc)} &
  \multicolumn{1}{l}{(Myr)} &
  \multicolumn{1}{l}{(Myr)} &
  \multicolumn{1}{l}{objects} & 
  \multicolumn{1}{l}{(this work)}  &
   \multicolumn{1}{l}{(\%)}  \\[1ex]
\hline\\[-0.8ex]
AB Doradus &              ABD & 34$\pm$26        & 100 &     100 / 50 & 16 (16) & 5 & 31$^{+13}_{-9}$ \\[0.8ex]
Argus &                       ARG & 106$\pm$51     & 40  &       50 / 20 & 1  & 0 & - \\[0.8ex]
$\beta$--Pic MG &              BPC & 31$\pm$21        & 22  &      30 / 10  & 17 (17) & 5 & 29$^{+13}_{-8}$ \\[0.8ex]
Carina &                       CAR & 85$\pm$35       & 35  &      35 / 10  & 26 (24) & 5 & 21$^{+10}_{-5}$ \\[0.8ex] 
Columba &                   COL & 82$\pm$30       & 35  &      40 / 15 & 17 (16) & 6 & 38$^{+13}_{-10}$ \\[0.8ex]
$\epsilon$--Cha &        ECH & 108$\pm$9       & 9    &      9 / 6    & 10 (10) & 2 & 20$^{+17}_{-7}$  \\[0.8ex]
Octans &                      OCT & 141$\pm$34     & 38  &      38 / 10 & 11 (10) &  5 & 50$^{+14}_{-14}$ \\[0.8ex]  
Tuc--Hor &                    THA & 48$\pm$7          & 33 &      33 / 10  & 14 (14) & 3 & 21$^{+14}_{-5}$ \\[0.8ex] 
\hline\\[-0.8ex]
\end{tabular}
\label{tab:summary_associations}
\tablefoot{\tablefoottext{a}{Ages are derived from the convergence method described in \cite{Torres2006} using new radial velocity values, results to appear in Torres et al. in prep.}\tablefoottext{b}{Effective number used to calculate MF values shown in parentheses.}\tablefoottext{c}{Systems either confirmed from PM analysis or high probability ($\ge$95\% detections) based on single-epoch data}.
\tablefoottext{d}{In the separation range 3-1000\,au considering observed targets with sensitivities $\ge$95\% of our sample.}
}

\end{table*}}

All the targets are bright enough (V $<$ 16 mag or K $<$ 13 mag) to be used as AO reference stars. For the majority of the stars, we used the visible wave-front sensor (VIS WFS). For the reddest objects we used the IR WFS with the N90C10 dichroic, which sends 90\% of the light to NAOS and 10\% to the infra-red detector. Since all the sources are very bright in the near-IR, the observations with the VIS WFS were performed with intermediate- and narrow-band filters to avoid saturation of the primary star. The first selection of filters was IB\,2.21 ($\lambda_{0}$, $\Delta\lambda$=2.21, 0.06$\,\mu$m) and NB\,2.17 ($\lambda_{0}$, $\Delta\lambda$=2.17, 0.023$\,\mu$m). During the observations, the IB 2.21 filter showed strong optical ghosts that prevented us from detecting companions in some particular regions of the detector. Therefore, we replaced it with the IB\,2.27 ($\lambda_{0}$, $\Delta\lambda$=2.27, 0.06$\,\mu$m) filter. The observations with the IR WFS were performed with the broad band $\mathrm{K_s}$ filter ($\lambda_{0}$, $\Delta\lambda$=2.18, 0.35$\,\mu$m).
All the observations from 2006 and 2008 used the S27 objective, which provides a field of view (FOV) of 27.6\arcsec\,$\times$\,27.6\arcsec.  The data from 2011 and 2012 used the S13 objective, with a FOV of 14\arcsec\,$\times$\,14\arcsec. 


We observed each object using a jitter sequence to remove the sky contribution and to correct for bad pixels. The sky emission was estimated from the median image of all offsets positions. The average total exposure time for each source was $\approx$12 minutes.  We reduced the data using the recommended {\it eclipse} ESO software \citep{Devillard1997} and the {\it esorex} driven interface {\it gasgano}. Dark subtraction, bad pixel correction, and flat-field correction were applied to each frame prior to aligning and stacking.

To derive accurate astrometry of the potential companions, we derived the plate scale and orientation on the infrared detector, using archival observations of the astrometric field $\theta^{1}$ Ori C \citep{McCaughrean1994}, and the astrometric binaries IDS 2150 and IDS 22141 \citep{vanDessel1993}; see \cite{Chauvin2012} for details. The values measured in different epochs are displayed in Table~\ref{tab:true_north_calib}. For each target, we have adopted the plate scale and orientation closest to the date of observation.

A summary of the NACO observations is provided in Table~\ref{tab:data_summary}.

\subsection{Northern Targets: Lick data}

To observe the 40 northern targets in our sample, we used the adaptive optics facility at the Lick Observatory Shane 3\,m telescope \citep{Gavel2002} between 14--18 July 2008 and 8--9 July 2009. 

In 2008 the conditions were relatively good, although not photometric, with an average seeing between 0.8$\arcsec$ and 1.1$\arcsec$ in the visible, at the zenith. The targets were all bright enough to be used as a reference for the wave-front sensor.  All of the targets were observed using the $\mathrm{K_s}$ filter, which provides the best compromise between the AO performances and the sensitivity to faint and red companions. 

When a clear companion was detected in the images, an additional H band (and sometimes J band) image was acquired to provide colour information. All observations were performed using a dither pattern to efficiently remove cosmic rays, bad pixels and accurately compute the sky. The data were processed using {\it Eclipse}, including dark subtraction, flatfield correction, sky subtraction, frame registration, and stacking. A Hipparcos binary \citep[HIP\,66225,][]{Gontcharov2006} was observed every night to be used as an astrometric calibrator and derive an accurate plate scale and plate position angle on the sky (see Table~\ref{tab:true_north_calib}).

The 2009 observations aimed at obtaining second epoch images of the 21 targets with companion candidates detected in the first epoch images. Conditions were poorer, with a seeing of $\approx1. 5\arcsec$ and some cirrus. As we were most interested in deriving accurate astrometry, the observations were performed in a single band (using the $\mathrm{K_s}$ filter). The same observing strategy (five point dither pattern) and processing steps were applied. The same astrometric reference was used to calibrate the plate scale and position angle of the camera (see Table~\ref{tab:true_north_calib}).  

A description of the Lick data used in this work is included in Table~\ref{tab:data_summary}.

\begin{table}
\caption{Detector plate scales ($s\pm\sigma_\mathrm{s}$) and true north orientations ($t\pm\sigma_\mathrm{t}$)\label{pscale}}
\begin{tabular}{p{1.6cm} p{1.9cm} p{1.9cm} p{1.6cm}}
\hline\hline\\[-1ex]
Obs. date & Plate scale & True north\tablefootmark{a} & Target\\[0.3ex]
     & (mas\,/\,pix) & (deg) \\      \hline\\[-1.3ex]
\multicolumn{4}{c}{NACO}\\ \hline\\[-1.2ex]
24/04/2006  & 27.01$\pm$0.05 & -0.08$\pm$0.14 & $\theta^1$  Ori C\\
26/08/2006  & 27.02$\pm$0.05 & -0.12$\pm$0.20 & $\theta^1$  Ori C\\
06/04/2008  & 27.01$\pm$0.05 & -0.36$\pm$0.14 & $\theta^1$  Ori C\\ 
10/06/2008  & 27.01$\pm$0.05 & -0.39$\pm$0.15 & IDS\,2150, \\ 
       & & & $\theta^1$  Ori C\\
15/11/2011  &  13.27$\pm$0.05 & -0.47$\pm$0.14 & $\theta^1$  Ori C \\
04/07/2012  &  13.27$\pm$0.05 & -0.78$\pm$0.20 & IDS\,22141 \\ \hline\\[-1.4ex]
\multicolumn{4}{c}{Lick}\\ \hline\\[-1.3ex]
15/07/2008  & 75.7$\pm$0.1   &   +1.80$\pm$0.30 & HIP\,66225 \\
09/07/2009  & 75.7$\pm$0.2   &  -0.30$\pm$0.20 & HIP\,66225 \\ \hline
\end{tabular}
\label{tab:true_north_calib}
\tablefoot{\tablefoottext{a}{This value is subtracted from the measured position angle (PA).}}
\end{table}

{\tiny
\begin{table}
\caption{A summary of all AO-imaging data used in this work.  The number of observations refers to the number of reduced images as opposed to separate objects, i.e., an object can be counted twice if it was re-observed.}
\begin{tabular}{p{1.4cm} p{1cm}  p{1.2cm}  p{0.8cm} p{0.8cm} p{1cm}}
\hline\hline\\
  \multicolumn{1}{l}{Instrument} &
  \multicolumn{1}{l}{FOV} &
  \multicolumn{1}{l}{Filter} &
  \multicolumn{1}{l}{$\lambda_0$} &
  \multicolumn{1}{l}{$\Delta\lambda$} &
  \multicolumn{1}{l}{No. of} \\
  \multicolumn{1}{l}{} &
  \multicolumn{1}{l}{(arcsec)} &
  \multicolumn{1}{l}{} &
  \multicolumn{1}{l}{($\mu$m)} &
  \multicolumn{1}{l}{($\mu$m)} &
  \multicolumn{1}{l}{Obs.}  \\[1ex]

\hline\\
NACO & 14 $\times$ 14  & $\mathrm{K_s}$ & 2.18 & 0.35 & 91\\
 & 28 $\times$ 28  &  $\mathrm{K_s}$& 2.18 & 0.35  & 53  \\
& 28 $\times$ 28  & IB 2.27 & 2.27 & 0.06  & 82 \\
& 28 $\times$ 28  & IB 2.21& 2.21  & 0.06 &11\\
& 28 $\times$ 28  & NB 2.17 & 2.17 & 0.02 & 12\\
Lick & 20 $\times$ 20  & $\mathrm{K_s}$ & 2.15 & 0.32  & 70 \\
& 20 $\times$ 20  & H & 1.66 & 0.30 & 14\\
& 20 $\times$ 20  & J & 1.24 & 0.27  & 2\\[1ex]
\hline
\end{tabular}
\label{tab:data_summary}
\end{table}}


\subsection{Archival data and literature search}

We queried the ESO archive for related NACO data for objects that were observed only once in our service campaigns.  This way, we could utilise a second epoch to determine if the detected companion candidates are co-moving.  In total 64 objects had multi-epoch observations.  In addition to this, we queried the CDS for object-type determinations and multiplicity flags for all 201 targets studied in this work, (see Table~\ref{tab:all_targets}).  For objects previously identified as multiple systems, we checked whether the companions' properties were within the parameter space of our observations, which in turn, tests the completeness of our work. As a result, we did not find any previously identified systems in this parameter space which we failed to detect.

\subsection{Observational detection limits}
\label{detection_limits}

\begin{figure}[h]
\begin{center}
\includegraphics[width=0.46\textwidth]{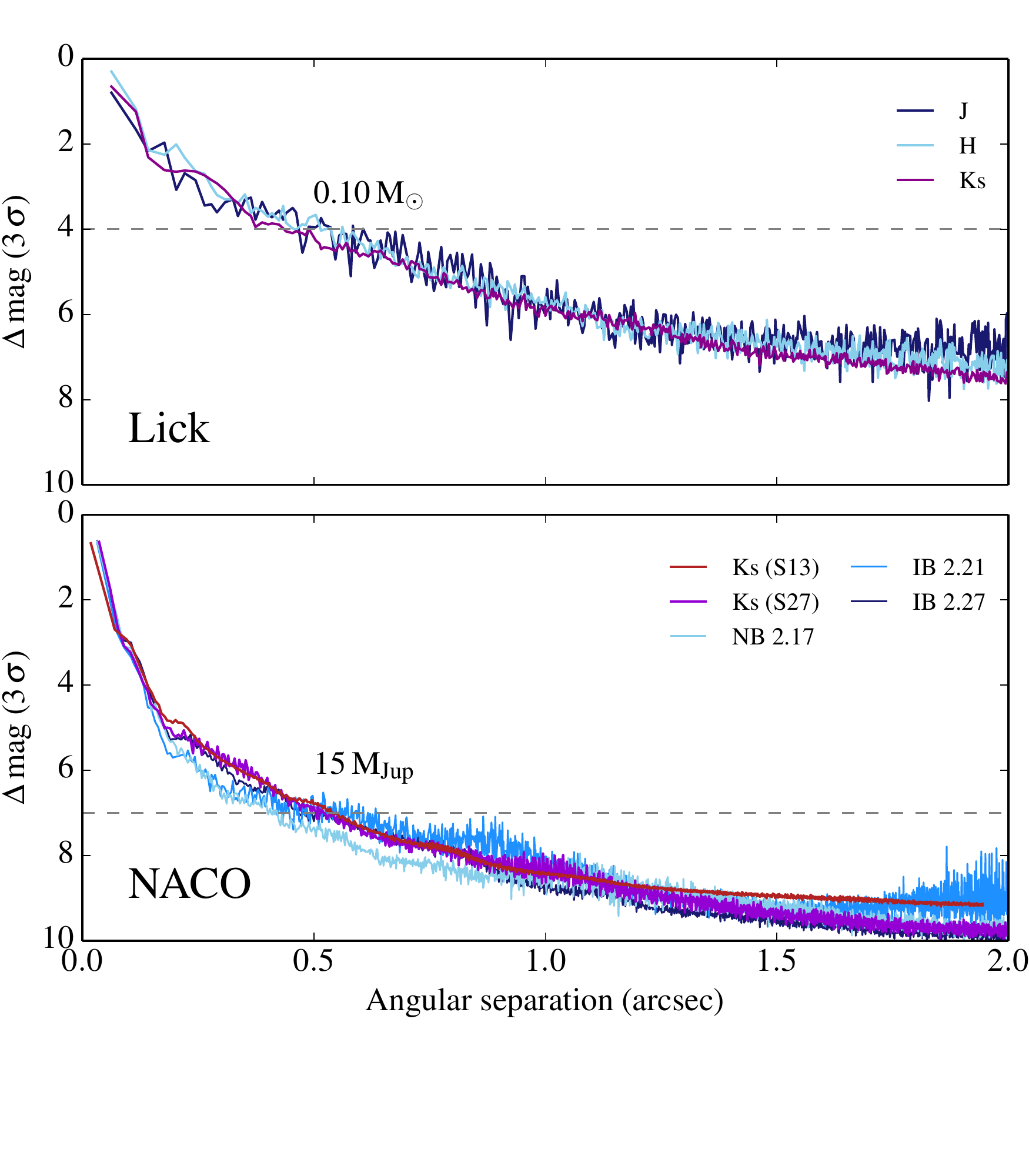}
\vspace{-0.5cm}
\caption{The average sensitivity limit in all filters for both NACO and Lick data. The limits were computed from the $3\sigma$ root-mean squared (RMS) of the PSF radial profile for each target, and then averaged within each filter.  The $\mathrm{K_s}$ filter for the NACO data is displayed for the S13 and S27 objective. The dashed lines represent our mass limits at 0\farcs5, according to \citet{Baraffe1998} evolutionary tracks, assuming an average age of 30\,Myr and a primary mass of 1\,M$_{\odot}$.}
\label{fig:av_sensitivities}
\end{center}
\end{figure}

Figure~\ref{fig:av_sensitivities} shows the average contrast in the selected filters as a function of the angular separation from the primary star for both the NACO (using the S13 and S27 objectives) and Lick data.  On average, we reach $\Delta\mathrm{K_s}$ = 7\,mag and $\Delta\mathrm{K_s}$ = 4\,mag  at an angular separation of 
0.5$\arcsec$, for the NACO and Lick data, respectively.  For an age of 30\,Myr (the average age of the SACY associations) and a primary mass of 1.0\,M$_{\odot}$, these differences in magnitude correspond to masses of $\sim$15\,M$_{\mathrm{Jup}}$ and $\sim$0.10\,M$_{\odot}$ (dashed lines in Figure~\ref{fig:av_sensitivities}), using the evolutionary tracks from \cite{Baraffe1998}. 

\section{Identifying multiple systems}
\label{sec:Identifying multiple systems}

In order to identify multiple systems in our sample, we have followed the next procedure:
first, we have identified all the point sources within the FOV of each target.  Then, we have determined the astrometry and photometry of these sources, and used these parameters to determine whether the identified point sources are bound companions or background objects. We provide all the details in the next subsections.

\subsection{Initial source detection}

Combining the NACO data and Lick data, we have imaged a total of 201 young targets.  In order to identify companions in our images, we first need to identify all possible point sources in the FOV. We used the source extraction software {\sc SExtractor} \citep{Bertin1996} to perform the source identification. Only detections of sources with 5 or more pixels with values $\ge 3\sigma$ were considered (where $\sigma$ is the standard deviation of the background signal).

\begin{figure}[h]
\begin{center}
\includegraphics[width=0.45\textwidth]{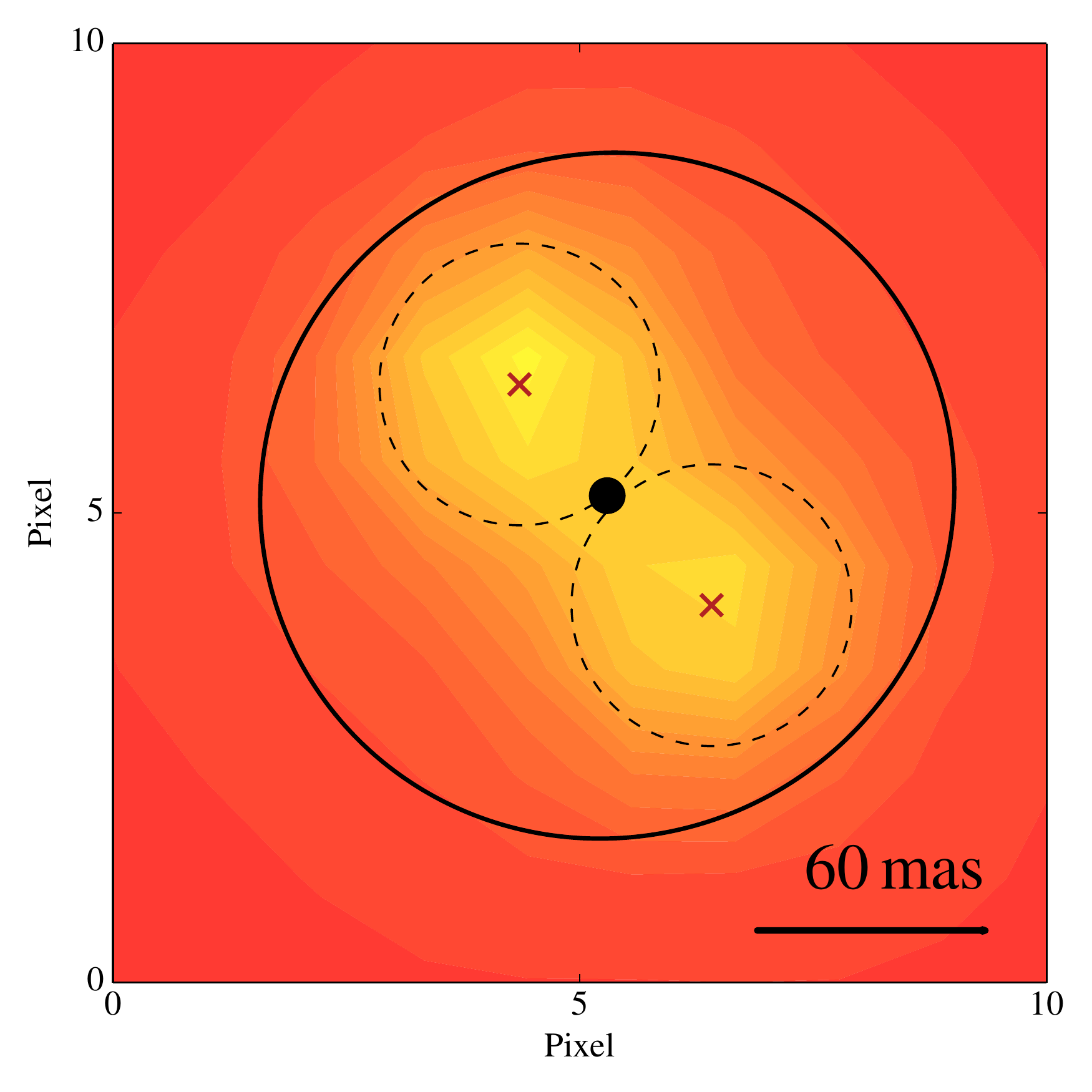}
\vspace{-0.5cm}
\caption{A section of the NACO image for target CD-30 3394B, showing two partially resolved components.  The black solid-lined ellipse is the aperature applied by {\sc SExtractor} and the black filled circle the centre of the ellipse.  The red crosses and black dashed circles indicate the centre and FWHM of the model PSF extracted from the primary CD-30 3394A, using {\sc PSFEx} for each source.}
\label{fig:blended_example}
\end{center}
\end{figure}

We visually checked each image to (i) confirm whether the detection was from a true point source as opposed to, for example, an optical ghost and (ii) search for potential severely blended binaries (with component separations $\sim$4\,pixel) that had been detected as one source.
With the catalogues of point sources for each observation we then modelled the PSFs using the software {\sc PSFEx} in conjunction with {\sc SExtractor}.  For the majority of observations, the primary targets were used to model the PSF. In the case that the primary target was resolved into a close binary, either further components in the FOV (as close to the centre of the image as possible) were used, or another bright target observed during the same night with the same instrumental setup and as close in time as possible.  Figure~\ref{fig:blended_example} shows a typical example of a blended object originally detected as a single source.  The center and the photometric aperture as derived by 
{\sc SExtractor} are represented by  the black circle and black ellipse, respectively.  Through the modelling of the PSF we were able to extract the astrometry and photometry for the two components, whose centres and FWHMs are shown by the red crosses and dashed circles, respectively.

\subsection{Relative photometry}

Our observations did not include photometric calibrator fields and, therefore, we can only perform relative photometry in our images.  However, the vast majority of our targets were imaged in the $\mathrm{K_s}$ band, so we can make use of the available 2MASS $\mathrm{K_s}$ values \citep{Skrutskie2006}.  In the cases where we discovered multiple components within the original PSF of 2MASS, we weighted the magnitudes according to the flux ratios of the components. These values are provided in the fifth column of  Table~\ref{tab:all_sacy_observed_targets}, as K(M$_1$), and have been used to compute the mass ratio of the multiple systems using evolutionary tracks \citep{Baraffe1998}, and the age estimation from Table~\ref{tab:summary_associations}.  

\subsection{Relative astrometry}

For every point source detected in each observation, we calculated the angular separation from the primary in detector coordinates and then applied the appropriate pixel scale and true north correction value (Table~\ref{tab:true_north_calib}).  
In the case of non-blended sources, we have used the parameters from the aperture extraction, while for severely blended sources we have 
used the ones coming from PSF modelling. We have performed both types of extractions (aperture and PSF) on a small number of targets and confirmed that, for non-blended sources, the resultant photometry and astrometry is consistent within the uncertainties.  We then have employed a different method to determine whether the detected point sources are bound companions or not, depending on the availability of multiple epochs.

\subsubsection{Uncertainties}

The main source of uncertainty comes from the temporal and spatial variability of the atmospheric conditions.  The uncertainties for the angular separation, $\rho(x,y)\pm\sigma_\rho$, and position angle, $\theta(x,y)\pm\sigma_\theta$, were derived using equations~\ref{eq:ang_sep} and \ref{eq:pa}, using the appropriate plate-scale and true north correction values. Table~\ref{tab:all_individ_detections} shows the parameters and their uncertainties of the multiple systems identified for each observation.

\begin{equation}
\label{eq:ang_sep}
\sigma_\rho = \Bigg[\bigg(\frac{\partial \rho}{\partial x}\sigma_\mathrm{x}\bigg)^2+\bigg(\frac{\partial \rho}{\partial y}\sigma_\mathrm{y}\bigg)^2\Bigg]^{\frac{1}{2}}
\end{equation}


where $\sigma_\mathrm{x}=x\,\sigma_\mathrm{s}$, $\sigma_\mathrm{y}=y\,\sigma_\mathrm{s}$ and $\sigma_\mathrm{s}$ is the uncertainty in the plate scale.

\begin{equation}
\label{eq:pa}
\sigma_\theta = \Bigg[\bigg(\frac{\partial \theta}{\partial x}\sigma_\mathrm{x}\bigg)^2+\bigg(\frac{\partial \theta}{\partial y}\sigma_\mathrm{y}\bigg)^2+\sigma_{\mathrm{t}}^2\Bigg]^{\frac{1}{2}}
\end{equation}

where $\sigma_t$ is the uncertainty in the true north correction.

\subsubsection{Multi-epoch data}
\label{sub_sec:multi}
\begin{figure}[h]
\begin{center}
\includegraphics[width=0.45\textwidth]{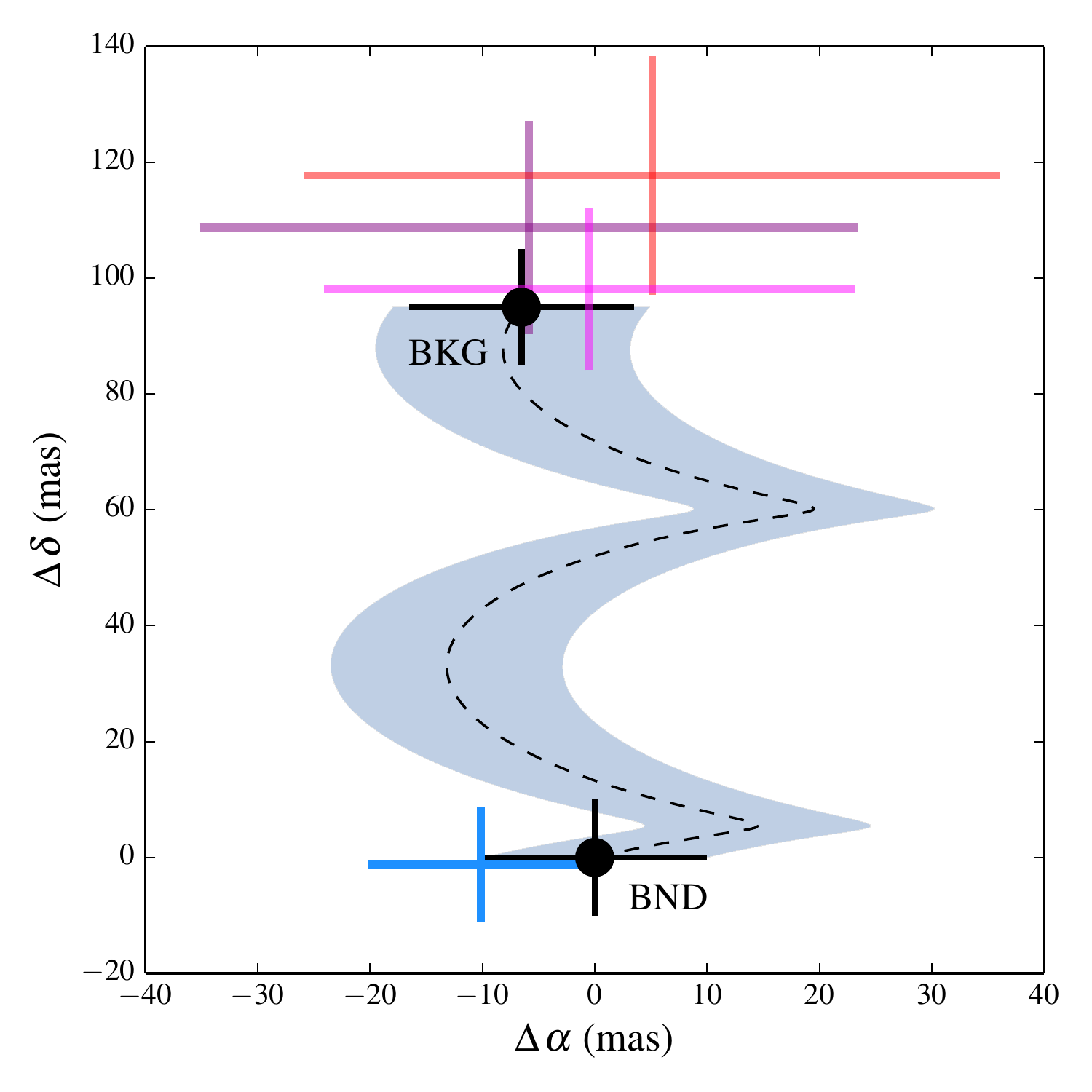}
\vspace{-0.5cm}
\caption{The relative movement of four point sources, in RA and DEC, for two observations of GSC 08350-01924.  The black dotted line and black markers represent the motion one would expect from a background (BKG) source with respect to the primary, and the shaded area represents the error space.  The dark blue marker represents a source showing no significant relative motion, consistent with a bound (BND) companion at 0.3$\arcsec$. The other coloured markers represent three background sources detected at separations of 10.5, 13.2 and 14.2$\arcsec$ from the primary.}
\label{fig:cd2211432_companions}
\end{center}
\end{figure}

To assess whether a pair of point sources are bound, and not a projection effect, we computed their relative motion and compared it to the motion one would expect for a background object.  The PM values were taken from UCAC4 \citep{Zacharias2012}. We then combined this with the parallactic motion, and produced the total relative motion one would expect to see if the point source were a background, stationary object \citep[see][]{Chauvin2005}.  An example is shown in Figure~\ref{fig:cd2211432_companions}.  We would expect a bound companion to show negligible movement with respect to the primary between our  multi-epoch observations, and this is our criterion for binary identification.  In those cases in which the PM of the source did not produce significant motion, given the time difference between observations, we used the criterion described in Section~\ref{sec:single_epoch}. 

\subsubsection{Single-epoch data}
\label{sec:single_epoch}

\newcommand{\BKG}{\textnormal{\tiny \textsc{bkg}}}
\newcommand{\BND}{\textnormal{\tiny \textsc{bnd}}}

For those targets with only one epoch of data (NACO data using S13 camera) we could use their properties, to provide statistical constraints on how likely is the source to be bound.  One way to do that is to estimate the potential contamination from background sources given the target's galactic co-ordinates, and the limiting magnitude of the observations.  However, galactic models such as \cite{Robin2003} have a limited resolution and, given our small FOV (13\arcsec $\times$ 13\arcsec), this method only provides a very rough estimate of the level of contamination.  To avoid this limitation and provide more realistic estimations, we decided to use our available multi-epoch data to assist in the classification of our single-epoch data.


We classified our own multi-epoch detections as either bound companions or background sources from PM analysis, see Section~\ref{sub_sec:multi}).  We then used two additional sources of data.  Firstly, \cite{Wahhaj2013}, a high-contrast direct imaging survey for giant planets.  We converted the H-band photometry into mass-ratios using the evolutionary tracks of \cite{Baraffe2003} and the ages quoted in the work.
Secondly, we used the data 
from \cite{Daemgen2014}, a multiplicity survey in Taurus conducted in K$_\mathrm{s}$-band.  From these two works, we only used sources classified by PM analysis (in total, 246 classified sources).

We then pseudo-randomly split this classified data: 2/3 as a training set, 1/3 a test set. Using support vector machines (SVM)\footnote{see Chapter 9 of \cite{Ivezic2014} for many applications of this technique in classification of astronomical objects such as differentiating variable and non-variable MS-stars using photometric colours.}, we determined the optimum soft-boundary in the physical separation-mass-ratio space that splits bound and background sources. On average, we reached a completeness and contamination rate of $>$80\% and $<$10\% respectively, when classifying our test set. With this classifier we then calculated the probability that our single-epoch sources are bound ($P_\mathrm{\BND}$) or background ($P_\mathrm{\BKG}$).  In this instance, there are only two labels: either bound or background and as a result,  $P_{\sc \mathrm{\BND}}=1-P_\mathrm{\BKG}$.  We repeated this procedure 1000 times and took the median result to account for the variations induced by different training and test datasets. The individual probabilities ($P_{\sc \mathrm{\BND}}$) of these companion candidates  with their respective properties are shown in Table~\ref{tab:binaries}.
 
To test whether this technique produces realistic probabilities, we calculated the bound probabilities for the 35 companions already confirmed by PM analysis.  The median bound probability was $\approx$99\%.

\begin{figure}[h]
\begin{center}
\includegraphics[width=0.49\textwidth]{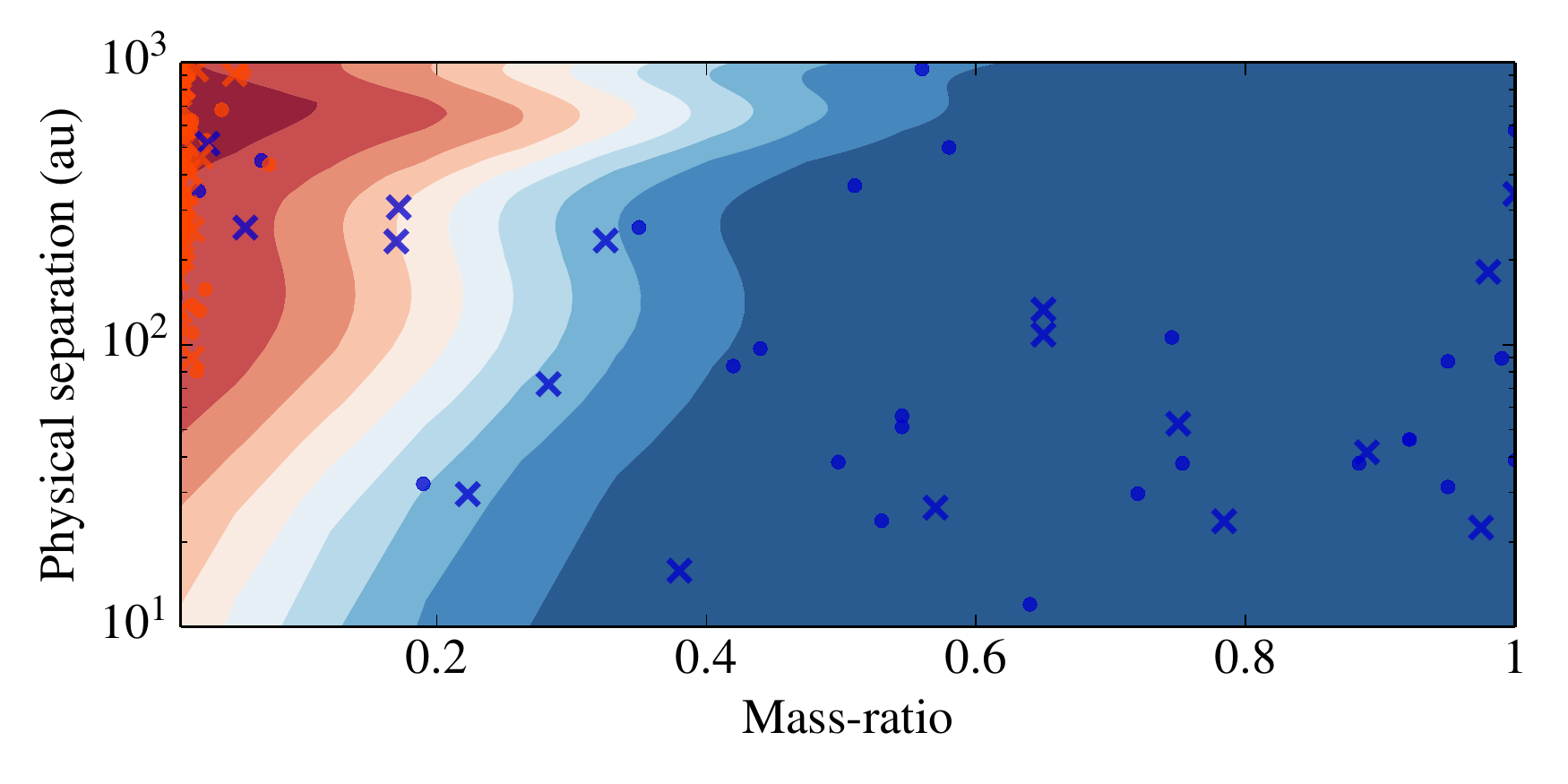}
\vspace{-0.5cm}
\caption{The probability of identifying a bound system from 100--0\% in intervals of 10\% from blue to red in physical separation--mass-ratio parameter space.  The circles and crosses represent the training and test set, respectively.  Blue markers represent bound companions, and red markers unbound companions.}
\label{fig:galactic_contam_2}
\end{center}
\end{figure}

Figure~\ref{fig:galactic_contam_2} shows the graphical result of this procedure. There is a clear trend towards high mass-ratio, closely-separated sources being bound  (as one would expect), and only very few bound sources lie in the parameter space mainly populated by background sources.

\section{Results}
\label{section:Results}

We have identified 31 potential multiple systems (28 binaries and 3 triples) from 113 confirmed SACY targets.  Of these 31 systems, 7 have been confirmed by PM analysis and 24 have a bound-probability $\geq$95\%\footnote{Based on this statistical analysis there is a 57\% probability that all 24 systems are bound from the product sum of their respective probabilities.} (see Section~\ref{sec:single_epoch}). 
The details of the observational properties from individual observations are displayed in Table~\ref{tab:all_individ_detections}.  A summary of the physical properties of the multiple systems is included in Table~\ref{tab:binaries}.


\begin{figure}[h]
\begin{center}
\includegraphics[width=0.45\textwidth]{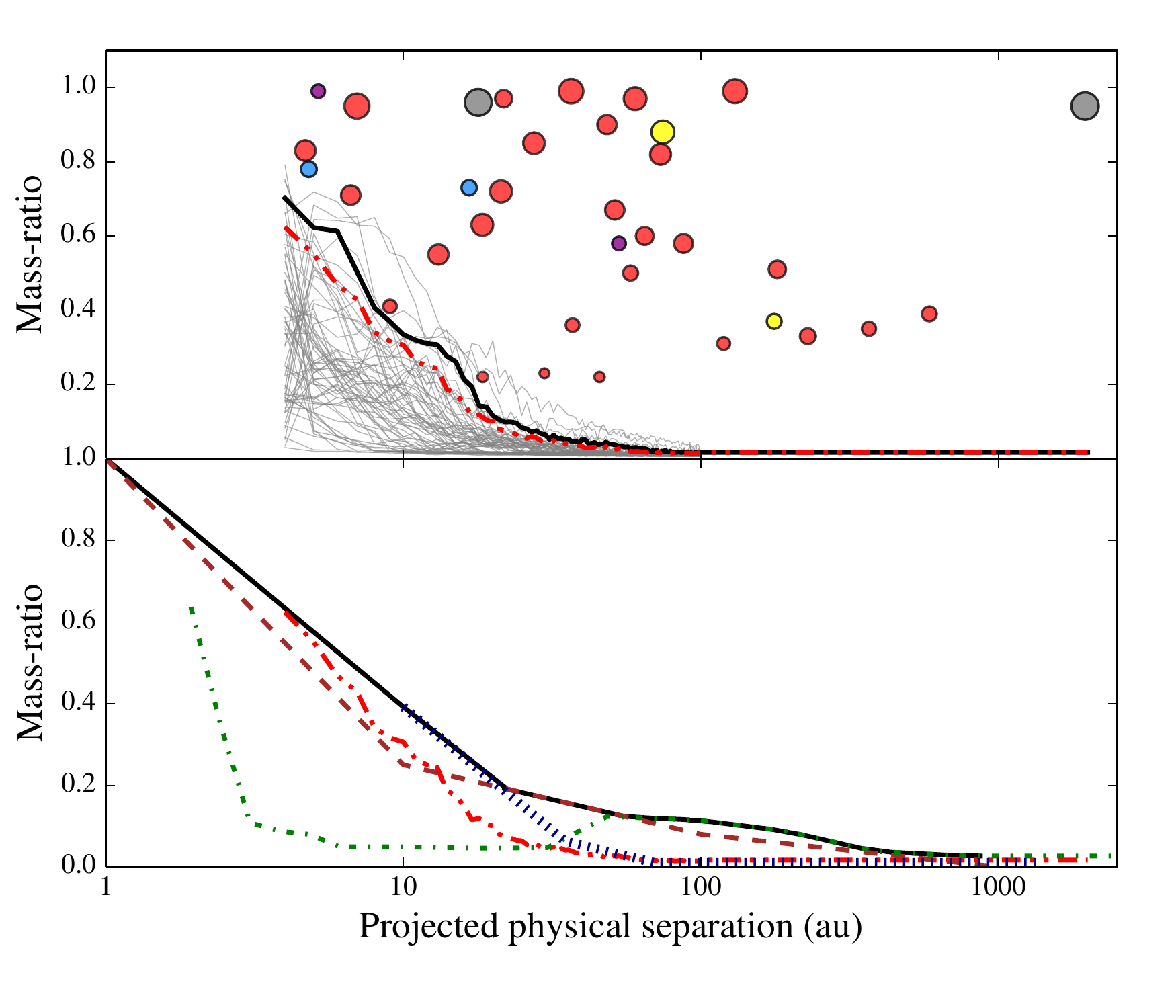}
\vspace{-0.5cm}
\caption{{\it Top panel:} The physical projected separations versus mass-ratio for identified multiple systems.  The filled red markers represent binary companions, the other coloured markers are components of triple systems with matching colours for the secondary and tertiary.  The size of the marker is proportional the companion's mass.  The grey lines represent the sensitivity limits for the individual targets. The
red and black lines represent the 90\% and 95\% detections limits for the whole sample, respectively. {\it Bottom Panel:} The 95\% detection limits for four studies, red: this work, blue: \cite{Daemgen2014}, green: \cite{Kraus2011} and brown: \cite{Raghavan2010}, used for analysis in Section~\ref{sec:Discussion}.}
\label{fig:ang_sep}
\end{center}
\end{figure}

We have followed the multiple component designations described in \cite{Tokovinin2005}, where systems are described by individual components and {\it super} components.  For example, a triple system composed of a binary system orbiting the primary (brightest) star would have the following designations: "A, B, *" and "Ba, Bb, B" following increasing hierarchy.  In such cases, there are two entries in Table~\ref{tab:binaries}, one to describe "A, Ba" and the other "Ba, Bb".

The analysis presented in the following subsections considers the 31 identified multiple systems from 8 SACY associations (outlined in Table~\ref{tab:summary_associations}). We note that have also identified 28 multiple systems in the sample of non-SACY members.  A summary of these systems' properties can be found in Appendix~\ref{sec:non_sacy_targets}.

We have defined two ranges of physical separations in our analysis, derived from the sample sensitivities displayed in Figure~\ref{fig:ang_sep}.  The first, 3-1000\,au, is the most extensive, considering 109 from the 113 observed targets, and including all the 34 companion detections (28 binaries and 3 triples).  We define another as 10-1000\,au, that will allow us to compare our results with other studies.  For this parameter space, we again consider 109 observed targets but only 27 companions detections (25 binaries and 1 triple).

\subsection{Multiplicity frequencies}
\label{sec:mfs}

The frequency of multiple systems in a population can be computed in a number of ways, see \cite{Reipurth1993}.  In this paper we consider the following quantities: multiplicity frequency (MF), companion star frequency (CSF) and triple frequency (TF).  

We have identified 31 multiple systems from observations of 113 confirmed SACY targets. Therefore, we estimate a raw multiplicity frequency (MF$_\mathrm{raw}$) of 27.4$^{+4.5}_{-3.8}$\%,  and a raw companion star frequency (CSF$_\mathrm{raw}$) of 30.1$^{+4.6}_{-3.9}$\%.  However, as mentioned previously, these frequencies include companions that would not be detected around all of the targets due to different contrasts and distances (see Figure~\ref{fig:ang_sep}).   When we remove these targets, we derive a multiplicity frequency of MF$_{3-1000\,\mathrm{au}}$=28.4$^{+4.7}_{-3.9}$\% in the separation range 3-1000\,au. All of our companion detections are above 95\% sensitivity curve and therefore we do not disregard any of them in our calculation.

\subsubsection{Higher-order multiple systems}

Considering the higher-order systems separately (3 triple systems) we derive a triple frequency of TF$_{3-1000\,\mathrm{au}}$=2.8$^{+2.5}_{-0.8}$\% in the separation range of 3-1000\,au.  
It is interesting to note that 2 of the 3 triple systems identified in this work have secondary components at very small separations (4.8 and 5.2\,au), that are only detectable due to the small distance to the systems (31 and 17\,pc).

\subsubsection{Multiplicity frequency and primary mass}

\begin{figure}[h]
\begin{center}
\includegraphics[width=0.49\textwidth]{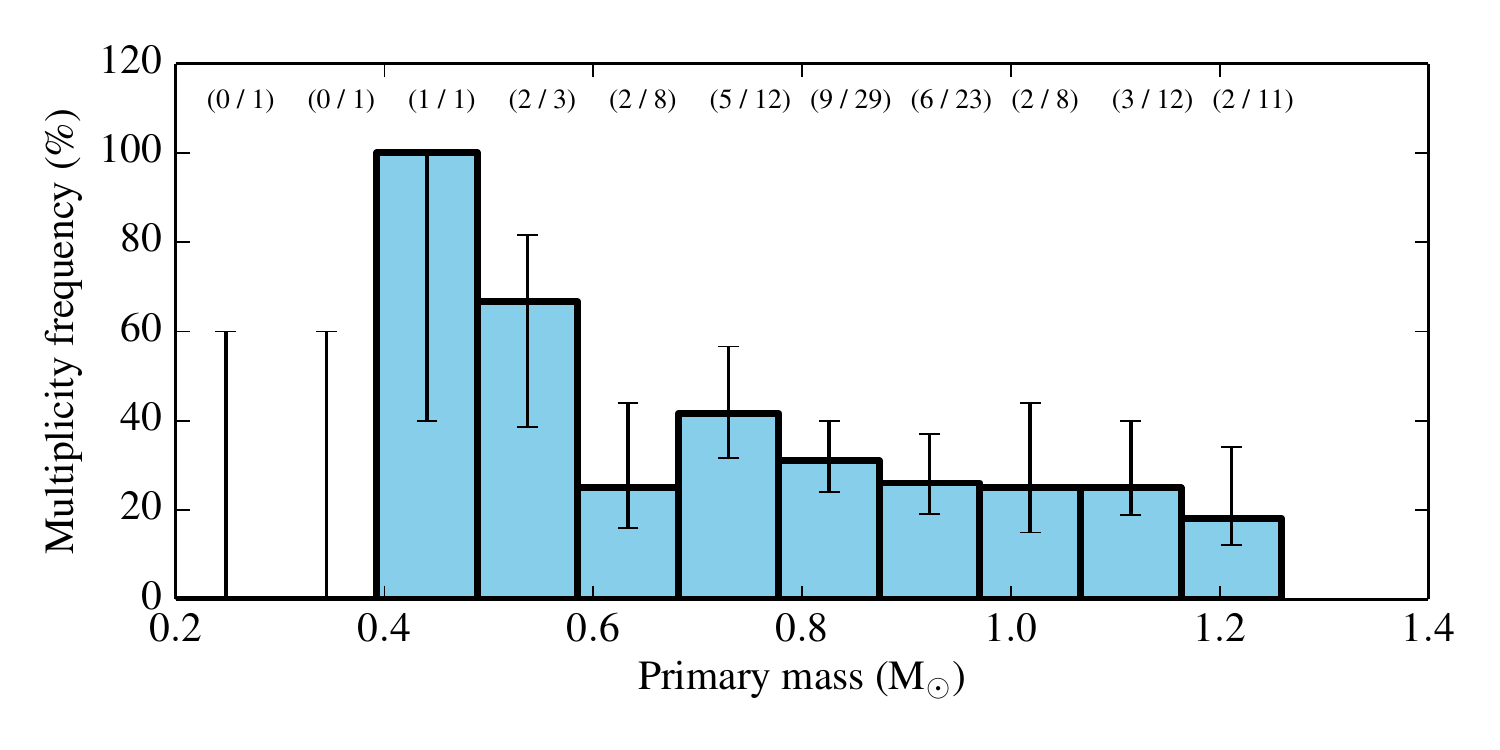}
\vspace{-0.5cm}
\caption{Primary mass versus multiplicity frequency for systems in the separation range 3-1000\,au. The data is binned according to the Freedman-Diaconis rule.  The fractions are shown by each mass bin.}
\label{fig:primary_mass_effect}
\end{center}
\end{figure}

We analysed the effect of the primary mass on the multiplicity frequency within our sample for the 109 systems in the separation range 3-1000\,au.  We binned our data according to the Freedman-Diaconis rule, which is suited for non-Gaussian distributions (bin width$\approx$0.1\,M$_\odot$).  The results are shown in Figure~\ref{fig:primary_mass_effect}.  The large uncertainties for masses of 0.25, 0.35 and 0.45\,M$_\odot$ are a result of the low-number fractions, 0/1, 0/1 and 1/1, respectively.

\subsection{Mass-ratios}
\label{subsec:m_rs}


We calculated the mass-ratios (q) of our multiple systems in two ways, to test the effect of using primary mass values derived with different methods. We note that, 
in the case of triple systems composed of a primary component A and an outer orbiting binary (Ba, Bb), the mass-ratio of the component Ba  
is derived as be M$_\mathrm{Ba}$/M$_\mathrm{A}$:\\[-2ex]

\begin{figure}[h]
\begin{center}
\includegraphics[width=0.49\textwidth]{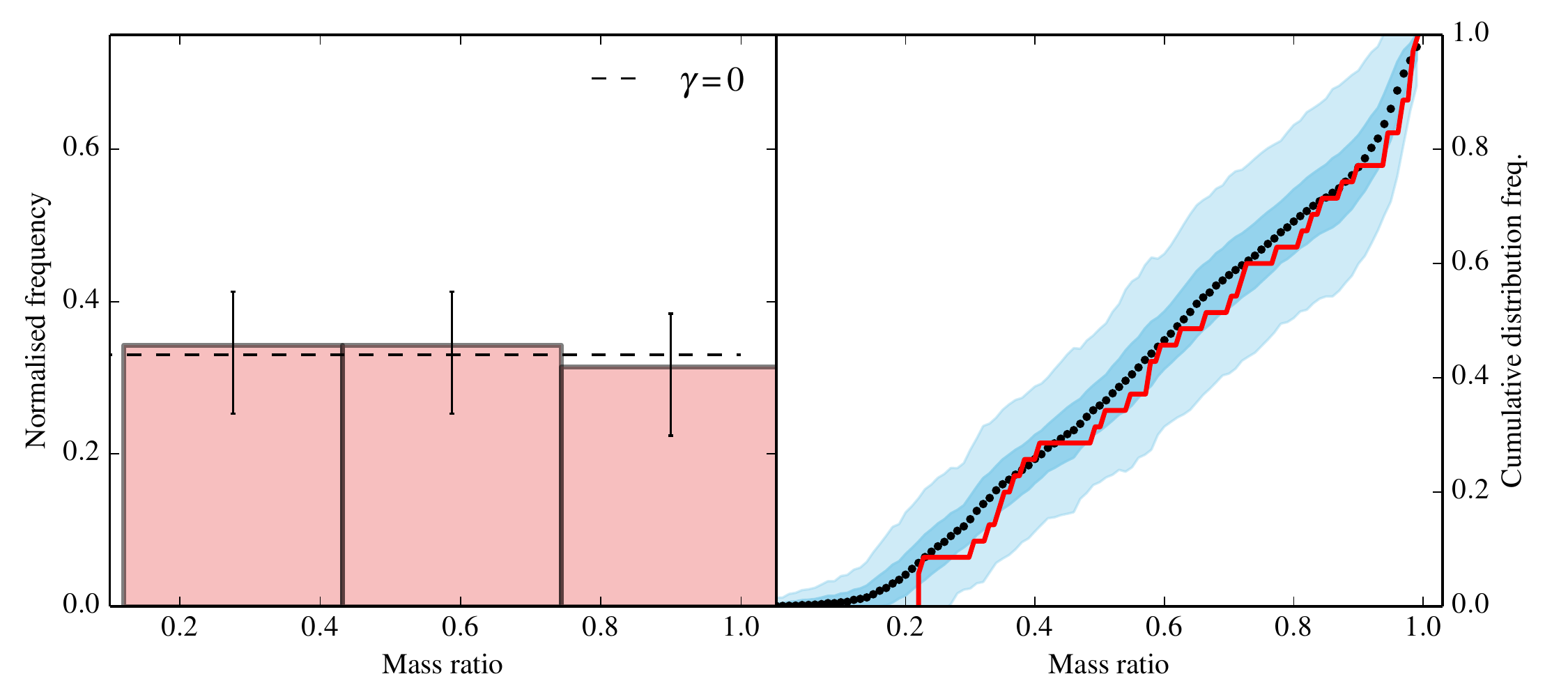}
\vspace{-0.5cm}
\caption{{\it Left panel}: Normalised frequency of mass-ratios.  The dashed line represents a power-law function with index $\gamma=0$.  {\it Right panel:} The cumulative frequency distribution of mass-ratios using the two methods outlined in Section~\ref{subsec:m_rs}.  Method (i) The red line represent the derived mass ratios.  Method (ii) The black dots, dark and light blue-coloured area represent the average, 1\,$\sigma$ and 3\,$\sigma$ variation in frequency from 100 realisations of the mass ratio, respectively.}
\label{fig:mass_ratio}
\end{center}
\end{figure}

(i) We derived primary mass values (M$_1$) from absolute magnitudes using kinematic distances ($D$) and ages with the evolutionary tracks of \cite{Baraffe1998}. We then calculated the companion masses from $\Delta$K values and computed the mass-ratios (the normalised cumulative distribution function (CDF) is shown by the red line in Figure~\ref{fig:mass_ratio}). \\[-2ex]

(ii) We used the $\Delta$K values of our components for a range of K$_1$ values, which is equivalent to a range of model-based M$_1$ values for the given age of the system 
\citep[0.8-1.2\,M$_\odot$, according to the evolutionary tracks of][]{Baraffe1998}. To obtain the uncertainty, we calculated the standard deviation of the mass-ratio as a function of the primary mass for a given system.  We then computed 100 mass-ratio distributions composed of pseudo-random realisations of each mass-ratio value and its uncertainty. We binned the data in 0.01 steps and took the average and standard deviation producing a smoothed, normalised CDF, shown by the black dots and shaded areas (dark and light: 1$\sigma$ and 3$\sigma$) of Figure~\ref{fig:mass_ratio}, respectively. 

The values and uncertainties used in further analysis (shown in Table~\ref{tab:binaries}) are obtained using method (ii) in order provide realistic uncertainties given that, at this time, the masses are model-based.

The mass distribution is usually described using a power law index, $\gamma$. We derive a value of $\gamma=-0.04\pm0.14$ using linear regression of 1000 pseudo-random realisations of the mass-ratio distribution considering their respective uncertainties.

\subsection{Physical separations}

We calculated the physical separations of the multiple systems using the kinematic distances shown in Table~\ref{tab:all_sacy_observed_targets}, and the average angular separation from our observations.  The top panel of Figure~\ref{fig:ang_sep} shows the physical separation versus mass-ratio for the 31 SACY multiple systems, considering both multiple-epoch and single-epoch observations with probabilities $\geq$95\%). The size of the markers represents the companion's mass.  In addition to this, the detection limits are shown for each multiple system (grey lines) and 90\% and 95\% detection limits for the entire observed sample in red and black, respectively.  This calculation demonstrates the powerful nature of the SACY dataset when observed with AO-imaging: we can probe very small physical projected separations across our sample down to low mass-ratios.

\begin{figure}[h]
\begin{center}
\includegraphics[width=0.49\textwidth]{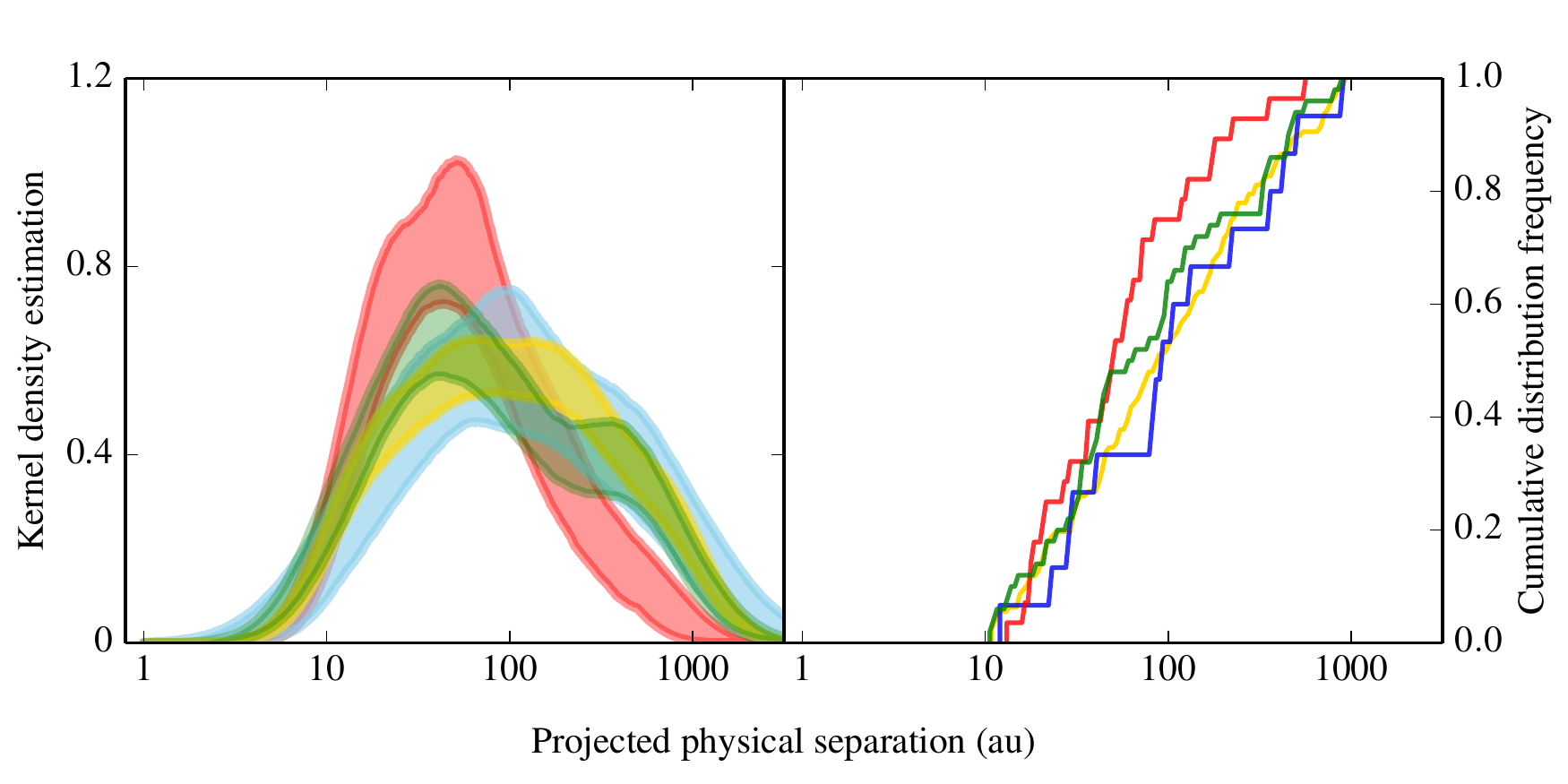}
\vspace{-0.5cm}
\caption{{\it Left:} Red, green, blue and gold shaded areas represent the 68\% confidence intervals for the kernel density estimation (KDE) for SACY, Taurus (K11), Taurus (D15) and the field (R10) respectively. The multiple systems considered are all above the master sensitivity curve defined in the bottom panel of Figure~\ref{fig:ang_sep} and in the separation range 10-1000\,au. {\it Right:} The cumulative distribution frequencies for the same systems.}
\label{fig:age_evol_phys_sep}
\end{center}
\end{figure}

\begin{figure}[h]
\begin{center}
\includegraphics[width=0.49\textwidth]{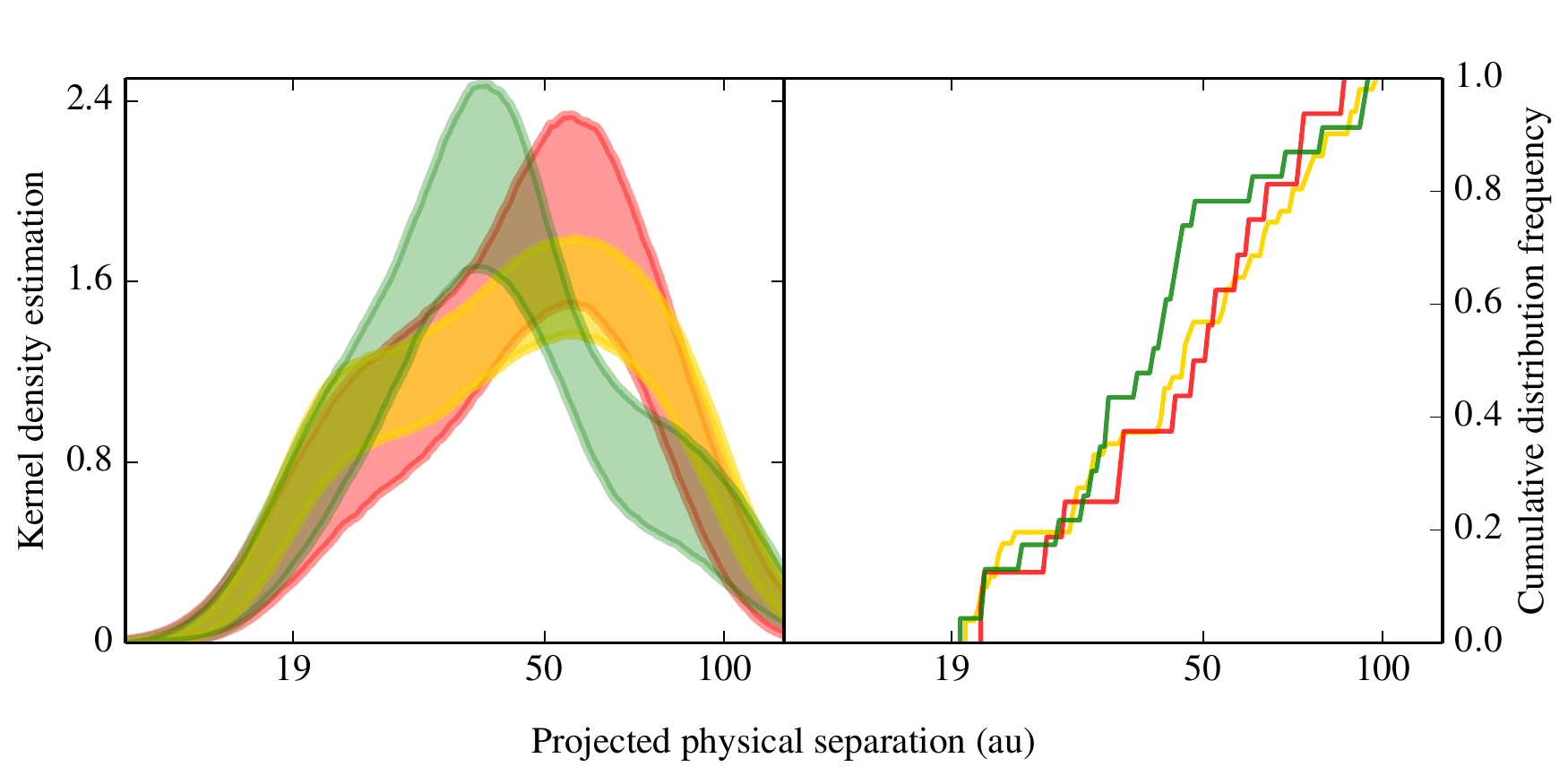}
\vspace{-0.5cm}
\caption{{\it Left:} Red, green and gold shaded areas represent the 68\% confidence intervals for the kernel density estimation (KDE) of multiple systems for SACY, Taurus (K11) and the field (R10), respectively.  The multiple systems considered are all above the master sensitivity curve defined in the bottom panel of Figure~\ref{fig:ang_sep} and in the separation range 19-100\,au. {\it Right:} The cumulative distribution frequencies for the same systems.}
\label{fig:sacy_king_field_comp}
\end{center}
\end{figure}

In addition to this, we generated separation distributions in two ranges. Firstly, 10-1000\,au,  for SACY, Taurus \citep[][hereafter K11 and D15]{Kraus2011, Daemgen2014} and the field \citep[hereafter R10]{Raghavan2010}, see Figure~\ref{fig:age_evol_phys_sep}. Secondly, 19-100\,au, for SACY, Taurus (only K11) and the field (see Figure~\ref{fig:sacy_king_field_comp}).  The first separation range was chosen to maximise the number of comparable systems due to different observational techniques. The second, to make the same comparison that was made in \cite{King2012b}.

First of all, we created 90\% sensitivity curves for each study in terms of mass-ratio and physical separation (shown by the coloured lines in the bottom panel of  Figure~\ref{fig:ang_sep}), using the available published data. In the case of R10, we used the sensitivity curve from Figure 11 of that paper. Then, we defined the master sensitivity curve as the upper boundary of all the individual curves (shown by the black solid line in the bottom panel of Figure~\ref{fig:ang_sep}).   We then only consider systems with parameters above this curve.

To perform this analysis we used kernel density estimations (KDEs), see \cite{scott2009multivariate} for details.  This technique minimises potential problems with the bin size and the phase. The results for the two separation ranges are shown in the left panels of Figures~\ref{fig:age_evol_phys_sep} and \ref{fig:sacy_king_field_comp}.  The plots shows the KDE from SACY in red, Taurus in blue and green, and the field in gold with 95\% confidence intervals (CIs) shown by the respective-coloured shaded areas. The CIs were calculated by considering the distribution of densities for each physical separation from 10$^5$ bootstrap iterations of the KDE. To see the differences between the distributions more clearly we have plotted them in the form of cumulative frequency distributions (CDFs) in the right panels of Figures~\ref{fig:age_evol_phys_sep} and \ref{fig:sacy_king_field_comp}.


The low-density nature of Taurus makes it very interesting to compare to our associations, as they most likely formed in a low-density environment too. We use both the work of K11 and D15 since both samples are complementary.  The statistical sample defined in D15 consists of 10 members from the extended Taurus region ($\sim$20\,Myr) and 5 from the young Taurus region ($\sim$2\,Myr), totalling 15 multiple systems.  That of K11 only considers the young region and in our defined parameter space, this produces a sample of 50 multiple systems for comparison.

The field represents a far more processed population of stars from a range of initial environments and, therefore, provides a mixed sample of evolved multiple systems.  The number of comparable systems in the field within our defined parameter range is 123.

Firstly, we tested the effect of primary mass on the separation distribution as the studies we are considering probe slightly different primary mass ranges.  

We performed a set of simulations using 10,000 synthetic binaries with primary masses between 0.2-1.4\,M$_\odot$.  We performed two sets of simulations which differed by their mass-ratio distribution.  In the first case, we considered the mass-ratio distribution as uniform across all primary masses. In the second simulation we adopt the information from Figure 5 of K11 as priors.  In the range M$_1$: 0.7-1.4\,M$_\odot$ the mass-ratio distribution as uniform. For the range M$_1$: 0.2-0.7\,M$_\odot$ there is a preference for more equal mass-ratio systems for lower primary masses.  We created a probability function for the secondary masses, and pseudo-randomly chose mass-ratio values from this distribution.  

The period values were drawn from the log-normal distribution defined in \cite{Raghavan2010}, $\mu$=5.03 and $\sigma$=2.28 log\,(day). We then fitted the separation distributions with log-normal distribution and evaluated their parameters. When comparing the two sets of simulations, we see that the peaks of the separation distributions are shifted to smaller separations for the lightest primary masses: mean values of 30\,au in the first case, and 60\,au in the second case. However, this difference is not significant given the separation range we are considering (10-1000\,au)  There is also no significant difference resulting from the mass-ratio distributions.  

Based on this result, we have compared separations distributions of field stars in the mass range of 0.7-1.4\,M$_\odot$, and SACY objects in the mass range of 0.2-1.2\,M$_\odot$.

Our comparison between the three populations is shown in Figure~\ref{fig:age_evol_phys_sep}. The shape of the SACY distribution differs from those of the other populations at a 1\,$\sigma$ level (68\% CIs are represented by the coloured filled areas in left panel of Figure~\ref{fig:age_evol_phys_sep}). The distribution departs in form from the other populations at $\approx$50-100\,au.  Potentially, this departure for wider systems is the result of a bias in the way we classify our objects as bound, favouring close-in systems, using the method described in Section~\ref{sec:single_epoch}.  However, to check this, we included companions regardless of their probabilities and the resultant distribution still differed from the other populations at a 1\,$\sigma$ level for the widest systems.

\section{Discussion}
\label{sec:Discussion}

The SACY sample offers a unique opportunity to study multiplicity among young stars down to very small physical separations.  Currently, it is not known whether all these associations shared a common origin, due to limitations on the accuracy of their galactic motion\footnote{The {\sc gaia} mission \citep{Lindegren2008} will provide extremely precise astrometry for members of such associations, with expected accuracies of 7-25\,$\mu$as.}. Whether these associations share a common origin or not, it is likely that these groups resulted from {\it sparse} star formation whereby fewer than 100 stars were formed in each group.  Therefore by studying these targets we are obtaining crucial information regarding this mode of star formation.  In this analysis we present the results assuming a common formation mechanism, i.e. grouping the data together in some cases from all associations and looking for relationships in physical parameters (i.e. primary mass, physical separation).


\subsection{Frequency of multiple systems}

The triple frequency is TF$_{3-1000\,\mathrm{au}}$=2.8$^{+2.5}_{-0.8}$\% in the separation range 3-1000\,au.  If we consider the range 10-1000\,au, we derive the value TF$_{10-1000\,\mathrm{au}}$=0.9$^{+2.0}_{-0.3}$\%. The value from \cite{Daemgen2014} in Taurus, TF$_{10-1000\,\mathrm{au}}$=1.8$^{+4.2}_{-1.5}$\%, is comparable within the uncertainties.

The multiplicity frequency for SACY is MF$_{3-1000\,\mathrm{au}}$=28.4$^{+4.7}_{-3.9}$\% in the separation range 3-1000\,au across the primary mass range 0.2-1.2\,M$_\odot$.   To compare to the work of \cite{Daemgen2014} in the Taurus SFR, we consider the separation range 10-1000\,au and derive a multiplicity frequency of MF$_{10-1000\,\mathrm{au}}$=25.7$^{+4.8}_{-3.7}$\%.  The multiplicity frequency for Taurus in that parameter space is MF$_{10-1000\,\mathrm{au}}$=26.3$^{+6.6}_{-4.9}$\%, so the two values agree very well within the uncertainties.  However, if we limit the primary mass of our targets the result is slightly different, as discussed below.

\begin{figure}[h]
\begin{center}
\includegraphics[width=0.49\textwidth]{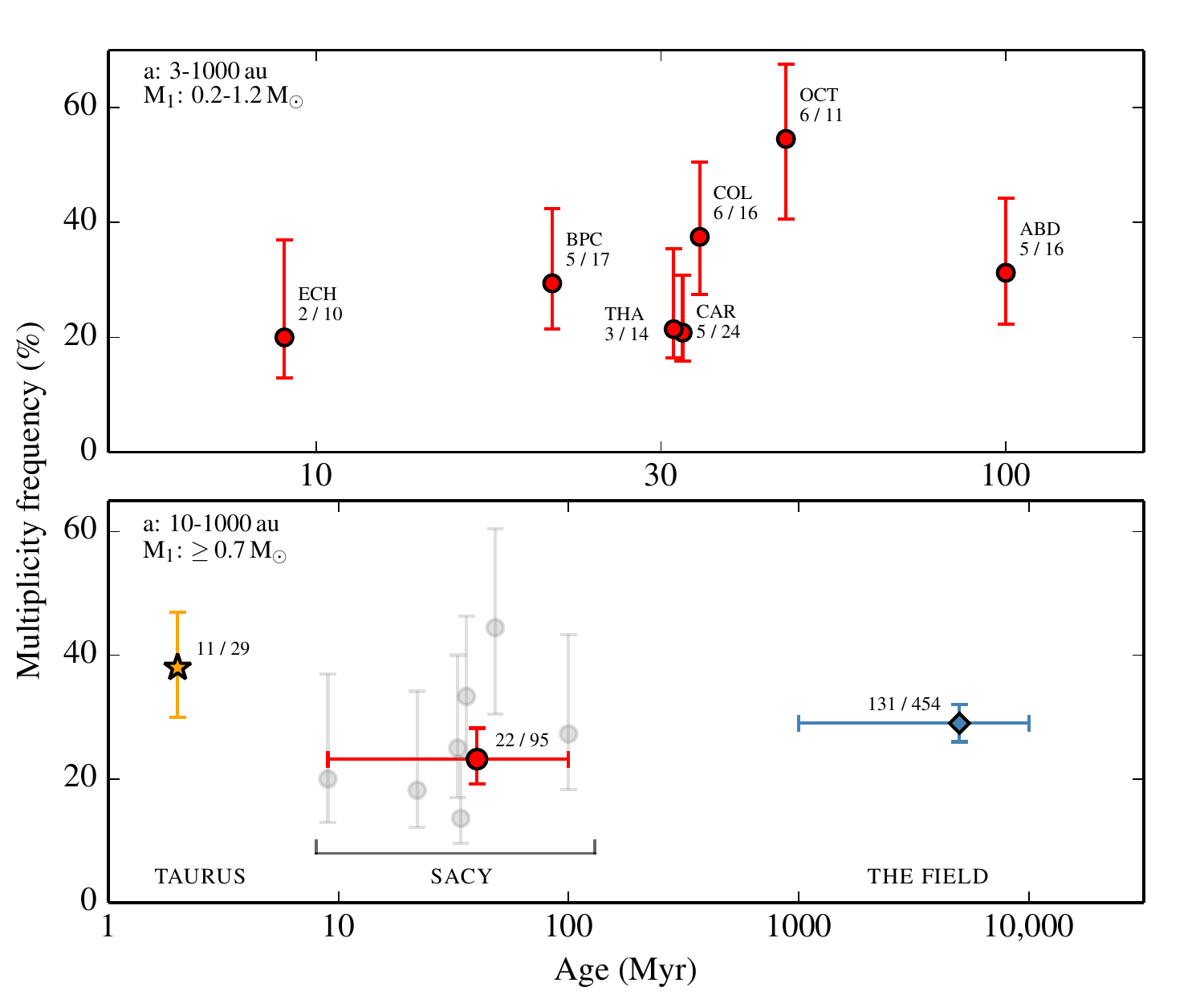}
\vspace{-0.5cm}
\caption{ Age versus multiplicity frequency.  {\it Top panel:} In the separation and primary mass range 3-1000\,au and 0.2-1.2\,M$_\odot$, respectively. {\it Bottom panel:} In the separation and primary mass range range 10-1000\,au and 0.7-1.4\,M$_\odot$, respectively. Data from \citep[][Taurus]{Daemgen2014} and \citep[][the field]{Raghavan2010} are shown by the star and diamond marker, respectively.  Argus (ARG) has been omitted as only 1 object was observed.}
\label{fig:mf_with_age}
\end{center}
\end{figure}

As shown by \cite{King2012a, King2012b}, collating studies from Taurus, $\rho$-Ophiuchus, Chamaeleon I, IC 348 and the ONC, and additionally the work of \cite{Kraus2011, Daemgen2014} studying Taurus, the multiplicity frequency of Taurus is marginally higher ($\approx$1\,$\sigma$) than that of other SFRs. In order to make a comparison between Taurus \citep{Daemgen2014}, SACY and the field \citep{Raghavan2010} we have selected targets in the separation range 10-1000\,au and the primary mass range $\ge$0.7\,M$_\odot$.  In the case of SACY the primary mass range is 0.7-1.2\,M$_\odot$ for Taurus and the field it is 0.7-1.4\,M$_\odot$. The resultant multiplicity frequencies (MF$_{10-1000\,\mathrm{au},\,\mathrm{M}_1\ge0.7}$) are shown in the bottom panel of Figure~\ref{fig:mf_with_age}.  The result of SACY (23.2$^{+4.8}_{-3.7}$\%) agrees with the field (28.9$^{+3.0}_{-2.9}$\%) and marginally disagrees with Taurus (37.9$^{+9.5}_{-7.9}$\%) at $\approx$1\,$\sigma$ level.  Based on these results, within this range, there is marginal evidence that the Taurus region has a higher frequency of multiple systems then SACY and the field. This result further supports the idea that Taurus has a higher MF value than other SFRs and associations. \cite{Kroupa2003} showed how environments such as Taurus are insufficient at binary disruption and, therefore, their high MF values are due to their binary populations being more pristine than that of heavily processed populations.  We would therefore expect, assuming similar primordial multiple system distributions, to see the same effect in the SACY associations. Currently, due to low numbers, we cannot say conclusively whether this property is shared or not.  There is no disagreement between SACY and the field, which, assuming common primordial multiple distributions, suggests that few multiple systems have been destroyed by dynamical processing in this range.

In contrast to the general trend of high multiplicity in low-density SFRs and associations, \cite{Brandeker2006} found an absence of multiple systems with separations $>$20\,au in the young \citep[6-8\,Myr,][]{Jilinski2005} cluster $\eta$ Chamaeleontis. This survey of 17 targets used NACO observations similar to those presented here, probing similar targets and mass-ratio systems. They found a probability $<$10$^{-4}$ that $\eta$ Chamaeleontis and TW-Hydrae share the same parent multiple system distribution.  Although $\eta$ Chamaeleontis is denser than the associations presented in this work, the authors conducted analysis to disregard dynamical processing as the cause of such a result, finding a collision time-scale 3000 times that of the cluster age.  The frequency of spectroscopic systems (SBs) in $\eta$ Chamaeleontis is statistically similar to that of the SACY associations (Riviere et al. submitted), $\approx$10\%.  The physical separations of these SBs are so tight that they should be free of dynamical processing and therefore similarities between regions support the idea of statistically similar binary populations.  However, the discs in $\eta$ Chamaeleontis are long-lived \citep{Sicilia2009}, considering its age, with respect to other young regions.  Therefore, at this time, it is still not clear whether the differences in multiplicity frequency between this cluster and other low-density SFRs and associations, 
are the result of dynamical processing or stem from environment-dependent star formation.  
 
However, since there is a general trend of high multiplicity frequency with low-density environments, it is likely that, at least in some part, it stems from the lack of dynamical processing compared to stars in populations such as the field \citep{Lada2003}.

\subsubsection{What is the effect of the primary mass on the frequency?}

It is common in many populations to observe an increase in multiplicity frequency with primary mass \citep[see Figure~12 of][]{Raghavan2010}.  We also noted the same relationship among the tightest  multiple systems, i.e. spectroscopic binaries \citep{Elliott2014}.  However, this is considering a very large mass range, from spectral types O-M.  We have investigated whether this relationship is also observable within our sample of stars (primary mass range 0.2-1.2\,M$_\odot$.)

Figure~\ref{fig:primary_mass_effect} shows the result of binning our data in terms of primary mass for the 109 targets with $\ge$95\% sensitivity from Figure~\ref{fig:ang_sep}).  There is no observed trend of increasing multiplicity frequency with primary mass.  There is a slight disagreement between some mass bins ($\approx$1\,$\sigma$) but only in the case of very low numbers ($\le$ 3).  The most reliable mass bins (with more than 10 objects) show no tendency of higher frequency with primary mass. This is in agreement with the results of \cite{Kraus2011} and \cite{Daemgen2014} within the same primary mass range.

\subsection{The impact of a flat mass-ratio distribution}

\cite{Duchene2013} compiled observations and discussed how random pairing from the IMF \citep{Chabrier2003} does not agree with the derived observational mass-ratio distributions for primary masses of $\sim$1-10\,M$_\odot$.  The mass-ratio power-law index is a much steeper function of the mass for lower-mass primaries ($\leq0.5$\,M$_\odot$).  

The mass-ratio distribution derived for the SACY sample (power-law index $\gamma$=$-0.04\pm0.14$), agrees with the compilation of observational results in \cite{Duchene2013}, in the primary mass range of $\approx$0.2-1.2\,M$_\odot$ (see left panel of Figure~\ref{fig:mass_ratio}). The derived $\gamma$ values in this range are very close to 0, i.e. flat distributions.   With our current observations we are not sensitive to more extreme mass-ratio systems (smaller than 0.2) at close-separations.  To investigate whether the power-law index continues to be flat in this primary mass range, we need to observe our sample with more sensitive techniques, such as extreme high-contrast imaging or sparse-aperture masking \citep{Evans2012}.  We aim to perform this analysis in future work.

\subsection{Is there dynamical evolution within the SACY sample?}
\label{sec:dynam_evol}

Another great advantage of the SACY sample is the range of PMS ages that we can probe ($\sim$10-100\,Myr). This allows us to search for any potential dynamical evolution within the sample on these time scales.

Figure~\ref{fig:mf_with_age} shows the multiplicity frequency for systems in the separation range 3-1000\,au.  If we consider the youngest and oldest associations, $\epsilon$-Cha (ECH) and AB Dor (ABD), we do not see significant differences in the multiplicity frequency.  Given the number of observations we have at this time, we are limited by low-number statistics.   To investigate this further, we need to observe more systems in each association to build up larger samples.  Given that there is only one association that does not agree with the other seven, we assume a statistically similar primordial multiple system distribution for all associations.

Although the parameters of such systems may evolve on a time-scale of 10-100\,Myr (see below) it is very unlikely that the systems will be entirely destroyed.  Most of the destruction of multiple systems occurs very early on, within a few crossing times of the clusters \cite[$<$1\,Myr, ][]{Parker2009} and, therefore, at the age of the associations studied here this process has long ended.

As mentioned above, we can also investigate the relationship between the projected physical separation and the age of the systems.  Although the frequency of systems has not significantly evolved, it is possible that we can observe significant differences in the system parameters.  To check this, 
we performed a KS test for three sets of ages: a) $<$30, b) 30-50, c) $>$50\,Myr.  The age groups were comprised of 7, 23, 6 companions, respectively.  The KS test produced one significant result; the distributions of b) and c) are very unlikely to be resultant from two different parental distributions (KS statistic: 0.18, p-value: 0.99). 
This statistical similarity implies there is no significant dynamical evolution of the physical separations on this time-scale (30-100\,Myr).  Due to the insignificant results from the other KS tests we cannot discern whether or not the other age-grouped samples agree or not.

{\tiny
\begin{table*}
\center
\caption{A summary of the properties of multiple systems identified in this work.}
\label{tab:binaries}
\begin{tabular}{p{3.8cm} p{1.2cm} p{0.7cm} p{0.7cm} p{0.7cm} p{0.7cm} p{0.9cm} p{0.7cm} p{0.7cm} p{0.7cm} p{0.7cm} p{0.7cm} p{0.7cm}}
\hline\hline\\[-1ex]
  \multicolumn{1}{l}{ID} &
  \multicolumn{1}{l}{Ang. sep.} &
  \multicolumn{1}{l}{PA} &
  \multicolumn{1}{l}{Dist.} &
  \multicolumn{1}{l}{Phys. sep} &
  \multicolumn{1}{l}{$\Delta$K} &
  \multicolumn{1}{l}{Comp.} &
  \multicolumn{1}{l}{Prob.\tablefoottext{a}} &
  \multicolumn{1}{l}{Pop.} &
  \multicolumn{1}{l}{$M_1$} &
  \multicolumn{1}{l}{$M_2$} &    
  \multicolumn{1}{l}{q} &
  \multicolumn{1}{l}{$\sigma_\mathrm{q}$} \\
  \multicolumn{1}{l}{} &
  \multicolumn{1}{l}{($\arcsec$)} &
  \multicolumn{1}{l}{(deg)} &
    \multicolumn{1}{l}{(pc)} &
  \multicolumn{1}{l}{(au)} &
  \multicolumn{1}{l}{(mag.)} &
  \multicolumn{1}{l}{} &
  \multicolumn{1}{l}{} &
  \multicolumn{1}{l}{} &
  \multicolumn{1}{l}{(M$_\odot$)} &    
  \multicolumn{1}{l}{(M$_\odot$)} &
  \multicolumn{1}{l}{} &
  \multicolumn{1}{l}{} \\[-0.1ex]
\hline\\[-1ex]
  GJ 4231                  & 0.16 & 298.4 & 30.6 & 4.8 & 0.58 & A, B   & .99 & ABD & 0.5 & 0.39 & 0.78 & 0.2\\
  GJ 4231                  & 0.54 & 0.4 & 30.6 & 16.6 & 0.70 & A, C   & .99 & ABD & 0.5 & 0.36 & 0.73 & 0.2\\
  TYC 91-82-1              & 0.74 & 110.4 & 87.6 & 64.8 & 1.67 & A, B   & .99 & ABD & 0.82 & 0.49 & 0.6 & 0.13\\
  BD-03  4778              & 2.53 & 208.5 & 71.5 & 180.9 & 2.19 & A, B   & - & ABD & 0.91 & 0.46 & 0.51 & 0.12\\
  Wolf   1225              & 1.44 & 232.7 & 15.1 & 21.7 & 0.07 & A, B   & - & ABD & 0.48 & 0.47 & 0.99 & 0.2\\
  HD 217379                & 2.24 & 239.0 & 32.7 & 73.2 & 0.63 & A, B   & .99 & ABD & 0.81 & 0.67 & 0.82 & 0.18\\
  HD 152555                & 3.79 & 58.0 & 46.8 & 177.4 & 3.89 & A, B   & .77 & ABD & 1.13 & 0.28 & 0.25 & 0.07\\
  GJ 3322                  & 1.36 & 149.5 & 37.8 & 51.4 & 0.78 & A, B   & .99 & BPC & 0.85 & 0.57 & 0.67 & 0.18\\
  GSC 08350-01924           & 0.73 & 19.9 & 66.3 & 48.4 & 0.17 & A, B   & - & BPC & 0.63 & 0.56 & 0.9 & 0.19\\
  2MASS  J05203182+0616115 & 0.42 & 235.0 & 71.0 & 29.8 & 2.19 & A, B   & .96 & BPC & 0.66 & 0.15 & 0.23 & 0.05\\
  CD-27  11535             & 0.08 & 278.8 & 87.3 & 7.0 & 0.17 & A, B   & - & BPC & 1.01 & 0.97 & 0.95 & 0.2\\
  CD-26  13904             & 0.27 & 81.8 & 78.9 & 21.3 & 0.93 & A, B   & - & BPC & 1.03 & 0.74 & 0.72 & 0.21\\
  2MASS  J09131689-5529032 & 0.14 & 128.7 & 131.7 & 18.4 & 1.48 & A, B   & .99 & CAR & 1.15 & 0.72 & 0.63 & 0.09\\
  CD-57  1709              & 5.84 & 220.5 & 100.5 & 586.9 & 2.28 & A, B   & .96 & CAR & 0.87 & 0.34 & 0.39 & 0.15\\
  2MASS  J08371096-5518105 & 5.72 & 23.5 & 180.8 & 1034.2 & 3.79 & A, B   & .38 & CAR & 1.11 & 0.18 & 0.16 & 0.04\\
  HD 22213                 & 1.67 & 281.1 & 52.4 & 87.5 & 1.87 & A, B   & .99 & CAR & 0.94 & 0.54 & 0.58 & 0.14\\
  BD-07  2388              & 0.11 & 328.6 & 42.6 & 4.7 & 0.62 & A, B   & .99 & CAR & 0.77 & 0.64 & 0.83 & 0.21\\
  CPD-62 1293              & 0.13 & 344.9 & 69.4 & 9.0 & 1.62 & A, B   & .98 & CAR & 0.7 & 0.28 & 0.41 & 0.14\\
  V* V1221 Tau             & 0.92 & 134.3 & 81.5 & 74.6 & 0.57 & A, B  & .99 & COL & 0.92 & 0.81 & 0.88 & 0.15\\
  V* V1221 Tau             & 2.17 & 320.7 & 81.5 & 176.6 & 2.5 & A, C   & .96 & COL & 0.92 & 0.34 & 0.37 & 0.11\\
  2MASS J08240598-6334024  & 0.17 & 288.2 & 118.6 & 20.7 & 3.141 & A, B   & .98 & COL & 1.2 & 0.34 & 0.29 & 0.05\\  
  2MASS J08240598-6334024  & 6.38 & 140.5 & 118.6 & 756.6 & 3.358 & A, C   & .29 & COL & 1.2 & 0.29 & 0.25 & 0.04\\
  2MASS  J04272050-4420393 & 0.43 & 179.9 & 86.2 & 37.1 & 2.15 & A, B   & .96 & COL & 0.81 & 0.29 & 0.36 & 0.15\\
  2MASS  J02303239-4342232 & 0.13 & 294.7 & 51.2 & 6.7 & 1.09 & A, B   & .99 & COL & 0.81 & 0.57 & 0.71 & 0.22\\
  BD-16  351               & 0.34 & 311.4 & 80.9 & 27.5 & 0.66 & A, B   & .99 & COL & 0.83 & 0.71 & 0.85 & 0.2\\
  CD-43  1395              & 0.26 & 324.6 & 141.0 & 36.7 & 0.02 & A, B   & .99 & COL & 0.93 & 0.92 & 0.99 & 0.16\\
  HD 272836                & 1.52 & 358.2 & 78.5 & 119.4 & 2.492 & A, B   & .96 & COL & 0.84 & 0.26 & 0.31 & 0.13\\
  HD 105923                & 1.96 & 145.1 & 116.9 & 229.1 & 2.81 & A, B   & - & ECH & 1.2 & 0.39 & 0.33 & 0.02\\
  2MASS  J12202177-7407393 & 0.23 & 3.7 & 114.8 & 26.4 & 3.91 & A, B   & .48 & ECH & 1.07 & 0.08 & 0.07 & 0.02\\
  TYC 9245-535-1            & 0.11 & 9.3 & 119.3 & 13.1 & 1.06 & A, B   & .99 & ECH & 1.14 & 0.63 & 0.55 & 0.1\\
  CD-30  3394A             & 12.06 & 116.7 & 162.4 & 1958.5 & 0.5 & A, Ba  & .82 & OCT & 1.2 & 1.14 & 0.95 & 0.08\\
  CD-30  3394A             & 0.11 & 43.2 & 162.4 & 17.9 & 0.18 & Ba, Bb & .99 & OCT & 1.14 & 1.10 & 0.96 & 0.14\\  
  HD 274576                & 3.14 & 15.8 & 117.1 & 367.7 & 2.51 & A, B   & .95 & OCT & 0.89 & 0.31 & 0.35 & 0.12\\
  2MASS  J04302731-4248466 & 0.49 & 193.4 & 122.8 & 60.2 & 0.13 & A, B   & .99 & OCT & 0.82 & 0.8 & 0.97 & 0.2\\
  BD-20  1111              & 4.01 & 301.0 & 129.7 & 520.1 & 3.56 & A, Ba  & .43 & OCT & 1.11 & 0.21 & 0.19 & 0.05\\
  BD-20  1111\tablefoottext{b}              & 0.07 & 340.0 & 129.7 & 9.1 & 0.46 & Ba, Bb & .98 & OCT & 0.21 & 0.18 & 0.86 & 0.14\\  
  CD-47  1999              & 0.22 & 321.5 & 166.7 & 36.7 & 3.56 & A, B   & .81 & OCT & 1.2 & 0.3 & 0.25 & 0.04\\
  2MASS  J06033540-4911256 & 0.75 & 90.0 & 173.8 & 130.4 & 0.06 & A, B   & .99 & OCT & 0.89 & 0.88 & 0.99 & 0.18\\
  BD-18  4452              & 3.18 & 190.7 & 16.7 & 53.1 & 0.81 & A, B  & - & OCT & 0.51 & 0.29 & 0.58 & 0.13\\  
  BD-18  4452              & 0.31 & 349.4 & 16.7 & 5.2 & 0.01 & Ba, Bb & - & OCT & 0.29 & 0.28 & 0.99 & 0.2\\
  TYC 8083-455-1            & 1.06 & 347.8 & 54.8 & 58.1 & 1.15 & A, B   & .98 & THA & 0.7 & 0.35 & 0.5 & 0.14\\
  2MASS  J05182904-3001321 & 0.69 & 76.5 & 66.2 & 45.7 & 2.48 & A, B   & .97 & THA & 0.76 & 0.17 & 0.22 & 0.07\\
  CD-44  1173              & 0.42 & 83.4 & 44.0 & 18.5 & 2.44 & A, B   & .96 & THA & 0.75 & 0.17 & 0.22 & 0.07\\
  HD 22705                 & 2.71 & 89.2 & 43.3 & 117.3 & 3.39 & A, B   & .52 & THA & 1.04 & 0.22 & 0.21 & 0.06\\

\hline\end{tabular}
\tablefoot{\tablefoottext{a}{Probabilities are shown unless system was confirmed by co-moving proper motion analysis.}\tablefoottext{b}{Ba, Bb has a probability $>$0.95, however, currently there is no information regarding its membership and is therefore not included in further analysis.}}
\end{table*}
}

\subsection{Comparing physical separation distributions}

The current understanding of binary formation invokes two formation channels; (i) fragmentation of the proto-stellar core, systems $\geq$100\,au \citep{Bodenheimer2001} (ii) via gravitational instability and fragmentation of the proto-stellar accretion disc of the primary star, systems $\leq$100\,au \citep{Toomre1964}.  If one of these two formation channels is dominant, one would expect to see a relative over- or under-abundance of systems between the two separation ranges, assuming that within our age range there is no significant evolution of the physical separations (see Section~\ref{sec:dynam_evol}).

Therefore, one interpretation of the over-abundance of multiple systems with separations $<$100\,au, a total of 21 systems, compared to wider systems, a total of 8, is that formation channel (ii) is dominant for multiple systems in our sample.

Another interpretation is that the original density of the SACY associations was much higher than the present day density.  This high-density environment would have dynamically processed many of the wider systems $>$100\,au early-on and, therefore, would lead to an observed relative under-abundance in this separation range.  However, this original density would have to be extremely high to account for such extreme processing.  In addition to this, the MF value of SACY is large and similar to that of Taurus which is an indication that the binary population has not been heavily processed.


\subsection{Universality of multiple systems}

Using a compilation of multiplicity surveys, \citet[][K12a,b hereafter]{King2012b, King2012a} claimed that binary (multiple system) formation is not the same everywhere, and that star formation is not a single universal process. This conclusion was partly based on the significant differences found in the CDF between young populations and field stars, in the separation range of 19-100\,au.  They claim that, in this particular separation range, these systems should not be dynamically processed and therefore their CDFs should be indistinguishable, independent of their age and initial environment.  However, \cite{Marks2014} dispute many of the results and claims of K12b.  They argue that if different initial densities are considered for the analysed SFRs, then the resultant multiple system properties are compatible with a single universal star formation process.  In fact, the separation range 19-100\,au is potentially not pristine, and in the case of Taurus, the comparable pristine range is only $\approx$20-30\,au.

For the sake of comparison, we have also produced a separation distribution in the range 19-100\,au, comparing 16 SACY systems, 23 Taurus systems, and 48 field systems. The results are shown in Figure~\ref{fig:sacy_king_field_comp}.  We note that we excluded data from D15, as in this separation range only seven systems were available for comparison. 

We find that the SACY and field distributions are statistically indistinguishable within the CIs.  We also performed a KS test which produced a p-value of 0.995, which is highly significant, this means it is very unlikely that the two distributions are realisations of different parent distributions.
If the separation range of 19-100\,au is considered a pristine range, this result is a strong indication that the stars born in the field and those born in associations form in a similar way.  

In the case of Taurus, the distribution shows a different shape, and has a p-value of 0.30 and 0.24 when compared to SACY and the field, respectively.
However, the p-values are not significant and therefore we cannot rule out that it is a realisation of the same parent distribution.

SACY and Taurus share many similar multiplicity properties (similar MF$_{10-1000\,\mathrm{au}}$ and TF$_{10-1000\,\mathrm{au}}$ values, a flat mass-ratio distribution). 

However, their respective separation distributions differ in both in the 10-1000\,au and the 19-100\,au range and additionally their MF$_{10-1000\,\mathrm{au},\,\mathrm{M}_1\ge0.7}$ are marginally discrepant.  If one population had been more heavily dynamically processed than the other, we would expect to see differences in both their MF values and their separation distributions.  We do see tentative differences in both quantities, implying that Taurus has experienced less dynamical processing than the SACY associations.


At this time, the significance of the differences between the populations is still low ($\approx$1\,$\sigma$), and therefore further data and analysis is still needed to clarify whether these differences are physical or merely an effect of low-number statistics.  One way to approach this problem is to focus on imaging a large number of targets in a pristine separation range.  However, calculating this pristine range can be problematic due to unknown initial densities.  Once again, the recently launched ESA mission, {\sc gaia}, will put important constraints on the galactic velocities of members with accurate astrometry. This will provide much better constraints on the initial densities of such associations.




\section{Conclusions}
\label{section:conclusions}

In this work we have presented multiplicity statistics of the SACY associations from AO-observations.  We have derived multiplicity properties of our sample from observations of 113 targets in the mass range 0.2-1.2\,M$_\odot$.  We compared these derived properties to other populations searching for statistical similarities and / or differences.

This study and previous work \citep{Elliott2014} form part of a larger project to characterise the multiplicity properties of the SACY associations across a large and continuous parameter space, using spectroscopy, sparse-aperture masking (SAM), speckle imaging, AO-imaging and classical direct imaging.  

Our observed sample consisted of 201 targets: 113 confirmed SACY members and 88 targets belonging to other populations (both from the field and PMS regions). As discussed in Section~\ref{sec:Observations and Data}, these 88 targets were not included in our analysis.  The parameters from individual observations can be found in Table~\ref{tab:all_individ_detections}, and a summary of the identified multiple systems in Table~\ref{tab:binaries_2}.

From the multiplicity study of the 113 confirmed SACY members our main conclusions can be summarised as follows:

\begin{enumerate}
\item We have identified 31 multiple systems (28 binaries, 3 triples) from observations of 113 targets.
\item Out of these 31 systems, 7 were identified from co-moving proper motion analysis, and 24 from contamination likelihood based on the component's parameters (with probabilities $\ge$0.95). 
\item The multiplicity frequency and triple frequency for SACY, in the separation range 3-1000\,au, are MF$_{3-1000\,\mathrm{au}}$=28.4$^{+4.7}_{-3.9}$\% and TF$_{3-1000\,\mathrm{au}}$=2.8$^{+2.5}_{-0.8}$\%, respectively.
\item In the separation range 10-1000\,au and for primary masses $\ge$0.7\,M$_\odot$ the multiplicity frequency (MF$_{10-1000\,\mathrm{au},\,\mathrm{M}_1\ge0.7}$) of SACY (23.2$^{+4.8}_{-3.7}$\%) is similar to the field (28.9$^{+3.0}_{-2,9}$\%) and discrepant at a 1\,$\sigma$ level to Taurus (37.9$^{+9.5}_{-7.9}$\%).
\item The multiplicity frequency is not a function of the primary mass, within the primary mass range of 0.2-1.2\,M$_\odot$. 
\item There is no evidence of dynamical destruction of multiple systems in the age range of 10-100\,Myr.
\item There is evidence that the multiple systems' physical separations do not evolve significantly on the time-scale 30-100\,Myr.
\item The mass-ratio distribution (power-law index $\gamma$=$-0.04\pm0.14$) is compatible with a uniform distribution ($\gamma$=0).
\item The SACY  and the field separation distributions are indistinguishable in the separation range 19-100\,au, suggesting a similar formation mechanism (if we assume this separation range is free of dynamical processing). 

\end{enumerate}

A detailed analysis of the statistical properties of the SACY associations in the context of star formation will be presented in a future paper.  It will include \cite{Elliott2014}, this paper, and \citep{Torres2008} in addition to collating all the available published information and data from the literature \citep[e.g.][]{Evans2012}. We can then compare a significant number of these young systems across a larger parameter space, and make our analysis much more comprehensive.  \\

\begin{acknowledgements}
The authors would like to thank the anonymous referee for the thorough work put into improving the content and direction of this manuscript. Thanks to Dr. Tokovinin for providing assistance and insightful discussions regarding the analysis of multiple systems in addition to tabular data.  Also thanks to Sebastian Daemgen for assistance with the multiplicity data of Taurus and to Simon Murphy for the discussion of stellar ages. This research has been funded by Spanish grants AYA2010-21161-C02-02 and AYA2012-38897-C02-01.  A. Bayo acknowledges financial support from the Proyecto Fondecyt de Iniciaci\'on 11140572.  This paper has made use of the SIMBAD database and VizieR catalogue access tool, CDS, Strasbourg, France.  This publication makes use of data products from the Two Micron All Sky Survey, which is a joint project of the University of Massachusetts and the Infra-red Processing and Analysis Center/California Institute of Technology, funded by the National Aeronautics and Space Administration and the National Science Foundation.  The following ESO observations were utilised in this work: 077.C-0483, 081.C-0825, 088.C-0506 and 089.C-0207.
\end{acknowledgements}

\bibliography{./biblio1}


%
%
%
\Online

\begin{appendix}

{
\onecolumn
\begin{landscape}
\section{All sources observed in this work}
\label{sec:appendix_a}
\tiny
\LTcapwidth=\textwidth
\begin{longtable}{p{3.5cm} p{1.2cm} p{1.2cm} p{1.2cm} p{1.2cm} p{1.2cm} p{1.2cm} p{1.2cm} p{1.2cm} p{2.0cm} p{1.5cm} p{1.2cm} p{1.2cm}}
\caption{All targets analysed in this work.} \label{tab:all_targets}\\
\hline\hline\\
  \multicolumn{1}{l}{ID} &
  \multicolumn{1}{l}{RA} &
  \multicolumn{1}{l}{DEC} &
  \multicolumn{1}{l}{V} &
  \multicolumn{1}{l}{K$_\mathrm{s}$} &
  \multicolumn{1}{l}{SpT} &
  \multicolumn{1}{l}{$\mu_\alpha$} &
  \multicolumn{1}{l}{$\mu_\delta$} &
  \multicolumn{1}{l}{Plx} &
  \multicolumn{1}{l}{Population} &
  \multicolumn{1}{l}{Simbad Class.} &
  \multicolumn{1}{l}{Lit. Flag\tablefootmark{a}} &
  \multicolumn{1}{l}{SACY Flag\tablefootmark{d}} \\[0.2ex]
  \multicolumn{1}{l}{} &
  \multicolumn{1}{l}{hh:mm:ss.s} &
  \multicolumn{1}{l}{dd:mm:ss} &
  \multicolumn{1}{l}{(mag.)} &
  \multicolumn{1}{l}{(mag.)} &
  \multicolumn{1}{l}{} &
  \multicolumn{1}{l}{(mas)} &
  \multicolumn{1}{l}{(mas)} &
  \multicolumn{1}{l}{(pc)} &
  \multicolumn{1}{l}{} &
  \multicolumn{1}{l}{} &
  \multicolumn{1}{l}{} &
  \multicolumn{1}{l}{} \\[1ex]  
\hline\\
\endfirsthead
\hline\hline\\
  \multicolumn{1}{l}{ID} &
  \multicolumn{1}{l}{RA} &
  \multicolumn{1}{l}{DEC} &
  \multicolumn{1}{l}{V mag.} &
  \multicolumn{1}{l}{K$_\mathrm{s}$ mag.} &
  \multicolumn{1}{l}{SpT} &
  \multicolumn{1}{l}{$\mu_\alpha$} &
  \multicolumn{1}{l}{$\mu_\delta$} &
  \multicolumn{1}{l}{Plx} &
  \multicolumn{1}{l}{Population} &
  \multicolumn{1}{l}{Simbad Class.} &
  \multicolumn{1}{l}{Lit. Flag\tablefootmark{a}} &
  \multicolumn{1}{l}{SACY Flag\tablefootmark{d}} \\[1ex]
\hline\\
\endhead
  HD 1405                & 00 18 20.9 &  +30 57 22 & 8.86 & 6.51 & K2V & 143.7 & -171.5 & \verb+~+  &  ABD & RS* & & \\
  HD 24681                & 03:55:20.4 & -01:43:45 & 9.05 & 7.25 & G5         & 46.2     & -90.2   & \verb+~+      & ABD                                       & *     &  & \\
  TYC 91-82-1             & 04:37:51.4 & +05:03:08 & 11.01 & 8.65 & \verb+~+          & 18.50    & -59.30  & \verb+~+      & ABD                                       & *     &  & AB         \\
  V* PX Vir               & 13:03:49.6 & -05:09:42 & 7.69 & 5.51 & G5V        & -191.13  & -218.73 & 46.10  & ABD                                       & BY*   &  & SB1        \\
  LP 745-70               & 16:33:41.6 & -09:33:11 & 11.24 & 7.55 & K9V     & -70.05   & -177.52 & 32.60  & ABD                                       & PM*   & A*F\tablefootmark{1}                           & \\
  HD 160934               & 17:38:39.6 & +61:14:16 & 10.45 & 6.81 & K7Ve       & -23.30   & 47.71   & 30.19  & ABD                                       & Fl*   & AB\tablefootmark{6},A*D, Aa,Ab\tablefootmark{1} & \\
  TYC 486-4943-1          & 19:33:03.7 & +03:45:39 & 11.15 & 8.66 & \verb+~+          & 18.0     & -65.5   & \verb+~+      & ABD                                       & *     &  & \\
  BD-03 4778              & 20:04:49.3 & -02:39:20 & 10.24 & 7.92 & \verb+~+          & 28.6     & -71.7   & \verb+~+      & ABD                                       & Ro*   &  & AB         \\
  HD 201919               & 21:13:05.2 & -17:29:12 & 10.64 & 7.58 & K6Ve       & 76.5     & -144.1  & \verb+~+      & ABD                                       & Ro*   &  & \\
  GJ 4231                 & 21:52:10.4 & +05:37:35 & 21.11 & 7.38 & M2.4V      & 119.17   & -150.29 & 32.79  & ABD                                       & Fl*   & AB, C & \\
  GJ 885 A                & 23:00:27.9 & -26:18:42 & 10.48 & 6.27 & (K7V)M+\verb+~+   & 113.2    & -169.2  & \verb+~+      & ABD                                       & *i*   &  & SB3        \\
  GJ 9809                 & 23:06:04.8 & +63:55:34 & 10.87 & 6.98 & M0.3V      & 171.46   & -58.55  & 40.81  & ABD                                       & Fl*   & A*Y,Za*m\tablefootmark{1}                      & \\
  CD-45 14955B            & 23:11:53.6 & -45:08:00 & 13.8 & 8.85 & M3Ve       & 85.2     & -84.7   & 19.7   & ABD                                       & *i*   &  & \\
  HD 222575               & 23:41:54.2 & -35:58:39 & 9.39 & 7.62 & G8V        & 69.49    & -67.53  & 15.70  & ABD                                       & Ro*   &  & \\
  HD 199058               & 20:54:21.0 & +09:02:23 & 8.62 & 6.96 & G5         & 37.2     & -56.9   & \verb+~+      & ABD\tablefootmark{10}                      & *     &  & \\
  TYC 1090-543-1          & 20:54:28.0 & +09:06:06 & 11.68 & 8.82 & \verb+~+          & 34.4     & -58.6   & \verb+~+      & ABD\tablefootmark{10}                      & V*    &  & \\
  HD 189285               & 19:59:24.1 & -04:32:06 & 9.43 & 7.84 & G5         & 14.2     & -55.8   & \verb+~+      & ABD\tablefootmark{10}                      & *     &  & \\
  V* PW And               & 00:18:20.8 & +30:57:22 & 8.81 & 6.39 & K2V        & 143.7    & -171.5  & \verb+~+      & ABD\tablefootmark{18}                      & RS*   &  & \\
  V* LO Peg               & 21:31:01.7 & +23:20:07 & 9.19 & 6.38 & K5-7V      & 133.38   & -145.24 & 40.32  & ABD\tablefootmark{18}                      & BY*   & AB,AC\tablefootmark{1}                         & \\
  BD+41 4749              & 23:19:39.5 & +42:15:09 & 8.93 & 7.22 & G0         & 77.52    & -66.90  & 19.94  & ABD\tablefootmark{18}                      & *     &  & \\
  HD 152555               & 16:54:08.1 & -04:20:24 & 7.82 & 6.36 & F8/G0V     & -37.25   & -114.05 & 21.40  & ABD                      & *     & AB & \\
  Wolf 1225               & 22:23:29.0 & +32:27:32 & 11.45 & 6.05 & M1V        & 255.3    & -207.8  & \verb+~+      & ABD, PMG\tablefootmark{12} & *i*   & AB\tablefootmark{3}                            & AB         \\
  CD-52 9381              & 20:07:23.7 & -51:47:27 & 10.5 & 7.39 & K6Ve       & 88.9     & -146.2  & \verb+~+      & ARG                                       & Ro*   &  & \\
  GJ 3322                 & 05:01:58.7 & +09:58:59 & 11.98 & 6.37 & M3V        & 12.09    & -74.41  & 30.12  & BPC                                       & Ro*   & AB\tablefootmark{4}                            & AB         \\
  2MASS J05200029+0613036 & 05:20:00.2 & +06:13:03 & 11.58 & 8.57 & K3         & 5.5      & -37.8   & \verb+~+      & BPC                                       & TT*   &  & \\
  2MASS J05203182+0616115 & 05:20:31.8 & +06:16:11 & 11.68 & 8.57 & K3         & 12.4     & -38.8   & \verb+~+      & BPC                                       & TT*   &  & AB         \\
  CD-27 11535             & 17:15:03.6 & -27:49:39 & 11.12 & 7.38 & K5Ve       & -9.30    & -48.30  & \verb+~+      & BPC                                       & Ro*   &  & AB, SB1    \\
  GSC 08350-01924         & 17:29:20.6 & -50:14:53 & 13.47 & 7.99 & M3Ve       & -5.0     & -54.7   & \verb+~+      & BPC                                       & *     &  & AB         \\
  CD-54 7336              & 17:29:55.0 & -54:15:48 & 9.55 & 7.36 & K1V        & -7.0     & -63.1   & \verb+~+      & BPC                                       & Ro*   &  & \\
  HD 161460               & 17:48:33.7 & -53:06:43 & 9.61 & 6.78 & K0IV       & -5.60    & -53.30  & \verb+~+      & BPC                                       & Ro*   &  & \\
  V* V4046 Sgr            & 18:14:10.4 & -32:47:34 & 10.94 & 7.25 & K5V        & 2.1      & -54.5   & \verb+~+      & BPC                                       & BY*   &  & SB2        \\
  GSC 07396-00759         & 18:14:22.0 & -32:46:10 & 12.78 & 8.54 & M1Ve       & 7.3      & -39.9   & \verb+~+      & BPC                                       & Em*   &  & \\
  Smethells 20            & 18:46:52.5 & -62:10:36 & 12.08 & 7.85 & M1Ve       & 13.7     & -82     & \verb+~+      & BPC                                       & pr*   &  & \\
  CD-31 16041             & 18:50:44.4 & -31:47:47 & 11.2 & 7.46 & K8Ve       & 10.6     & -77.8   & \verb+~+      & BPC                                       & Ro*   & A*J\tablefootmark{1}                           & \\
  TYC 6872-1011-1         & 18:58:04.1 & -29:53:04 & 11.78 & 8.02 & M0Ve       & 6.7      & -42.9   & \verb+~+      & BPC                                       & Ro*   &  & \\
  CD-26 13904             & 19:11:44.6 & -26:04:08 & 10.39 & 7.37 & K4V(e)     & 18.9     & -44.2   & \verb+~+      & BPC                                       & pr*   & AB\tablefootmark{2}                            & AB         \\
  CPD-72 2713             & 22:42:48.9 & -71:42:21 & 10.6 & 6.89 & K7Ve       & 94.1     & -54.4   & \verb+~+      & BPC                                       & Ro*   &  & \\
  V* WW PsA               & 22:44:57.9 & -33:15:01 & 12.07 & 6.93 & M4IVe      & 184.76   & -119.76 & 42.84  & BPC                                       & BY*   & A\tablefootmark{3}                             & \\
  V* TX PsA               & 22:45:00.0 & -33:15:25 & 13.36 & 7.79 & M5IVe      & 183      & -118    & \verb+~+      & BPC                                       & Fl*   & B\tablefootmark{3}                             & \\
  BD-13 6424              & 23:32:30.8 & -12:15:51 & 10.5 & 6.57 & M0Ve       & 139.20   & -83.40  & \verb+~+      & BPC                                       & Ro*   &  & \\
  HD 8813                 & 01:23:25.8 & -76:36:42 & 8.46 & 6.79 & G6V        & 100.16   & -20.93  & 21.52  & CAR                                       & pr*   &  & \\
  HD 22213                & 03:34:16.3 & -12:04:07 & 8.86 & 6.79 & G8V        & 80.7     & -35.4   & \verb+~+      & CAR                                       & Ro*   &  & AB         \\
  CD-44 1533              & 04:22:45.6 & -44:32:51 & 10.47 & 8.58 & K0V        & 32.0     & 2.8     & \verb+~+      & CAR                                       & Ro*   &  & \\
  CD-57 1709              & 07:21:23.7 & -57:20:37 & 10.73 & 8.7 & K0V        & -1.6     & 22.1    & \verb+~+      & CAR                                       & Ro*   &  & AB         \\
  TYC 8557-1251-1         & 07:55:31.6 & -54:36:50 & 11.44 & 9.19 & G9V        & -4.1     & 10.3    & \verb+~+      & CAR                                       & Ro*   &  & \\
  2MASS J08110934-5555563 & 08:11:09.3 & -55:55:56 & 11.52 & 9.4 & G8V(e)     & -3.5     & 15.1    & \verb+~+      & CAR                                       & pr*   &  & \\
  BD-07 2388              & 08:13:50.9 & -07:38:24 & 9.32 & 6.92 & K0         & -25.3    & -45.2   & \verb+~+      & CAR                                       & Ro*   &  & AB         \\
  2MASS J08371096-5518105 & 08:37:10.9 & -55:18:10 & 11.04 & 9.38 & G9V        & -7.4     & 18.0    & \verb+~+      & CAR                                       & pr*   &  &      AB?    \\
  CD-61 2010              & 08:42:00.4 & -62:18:26 & 10.95 & 8.83 & K0V        & -11.3    & 15.9    & \verb+~+      & CAR                                       & Ro*   &  & \\
  CD-53 2515              & 08:51:56.4 & -53:55:56 & 11.06 & 8.75 & G9V        & -11.6    & 12.7    & \verb+~+      & CAR                                       & Ro*   &  & \\
  2MASS J08521921-6004443 & 08:52:19.2 & -60:04:44 & 11.48 & 9.37 & K0V        & -2.8     & 12.0    & \verb+~+      & CAR                                       & pr*   &  &          \\
  2MASS J08563149-5700406 & 08:56:31.4 & -57:00:40 & 11.83 & 9.76 & K0V(e)     & -14.7    & 6.0     & \verb+~+      & CAR                                       & pr*   &  & \\
  TYC 8582-3040-1         & 08:57:45.6 & -54:08:36 & 11.71 & 9.35 & K2IV(e)    & -13.1    & 8.6     & \verb+~+      & CAR                                       & Ro*   &  & \\
  CD-49 4008              & 08:57:52.1 & -49:41:50 & 10.5 & 8.64 & G9V        & -23.00   & 15.40   & \verb+~+      & CAR                                       & Ro*   &  & \\
  CD-54 2499              & 08:59:28.7 & -54:46:49 & 10.08 & 8.4 & G5IV       & -18.3    & 15.2    & \verb+~+      & CAR                                       & Ro*   &  & \\
  CD-55 2543              & 09:09:29.3 & -55:38:27 & 10.21 & 8.4 & G8V        & -13.2    & 14.8    & \verb+~+      & CAR                                       & Ro*   &  & \\
  TYC 8174-1586-1         & 09:11:15.8 & -50:14:14 & 11.81 & 9.5 & K5Ve       & -19.45   & 7.42    & \verb+~+      & CAR                                       & Ro*   &  & \\
  2MASS J09131689-5529032 & 09:13:16.8 & -55:29:03 & 11.36 & 8.36 & G5V(e)     & -13.8    & 14.8    & \verb+~+      & CAR                                       & pr*   &  & AB         \\
  HD 309751               & 09:31:44.7 & -65:14:52 & 11.33 & 8.36 & G3V        & -12.40   & 19.00   & \verb+~+      & CAR                                       & pr*   &  & SB2        \\
  CPD-62 1293             & 09:43:08.8 & -63:13:04 & 10.44 & 8.6 & G6V        & -34.9    & 25.0    & \verb+~+      & CAR                                       & Ro*   &  & AB         \\
  TYC 9217-417-1          & 09:59:57.6 & -72:21:47 & 11.7 & 8.69 & K4Ve       & -24.2    & 29.0    & \verb+~+      & CAR                                       & Ro*   &  &          \\
  CD-69 783               & 10:41:23.0 & -69:40:43 & 10.27 & 8.38 & G8V(e)     & -30.0    & 24.4    & \verb+~+      & CAR                                       & Ro*   &  & \\
  TYC 8962-1747-1         & 11:08:07.9 & -63:41:47 & 12.07 & 8.29 & M0Ve       & -33.2    & 7.2     & \verb+~+      & CAR                                       & Ro*   &  & \\
  HD 107722               & 12:23:29.0 & -77:40:51 & 8.24 & 7.14 & F6V        & -65.1    & 10.8    & \verb+~+      & CAR                                       & TT*   &  & \\
  TYC 9486-927-1          & 21:25:27.4 & -81:38:27 & 11.85 & 7.34 & M1Ve       & 63.4     & -110.0  & \verb+~+      & CAR                                       & Ro*   &  & \\
  BD-03 5579              & 23:09:37.1 & -02:25:55 & 10.99 & 7.82 & K4Ve       & 57.5     & -47.4   & \verb+~+      & CAR                                       & Ro*   &  & \\
  CPD-53 1875             & 08:45:52.7 & -53:27:28 & 10.42 & 8.77 & G2V        & -13.8    & 7.7     & \verb+~+      & CAR\tablefootmark{10}                      & Ro*   &  & AB         \\
  CPD-55 1885             & 09:00:03.3 & -55:38:24 & 10.83 & 9.01 & G5V        & -12.7    & 16.4    & \verb+~+      & CAR\tablefootmark{10}                      & Ro*   &  & \\
  Ass Cha T 2-21          & 11:06:15.4 & -77:21:56 & 11.42 & 6.42 & G5Ve       & -16.5    & -0.5cm    & \verb+~+      & CHA I                                      & pr*   & D+w\tablefootmark{9}                           & \\
  2MASS J12571172-7640111 & 12:57:11.7 & -76:40:11 & 15.0 & 8.4 & M0e        & 10       & -32     & \verb+~+      & CHA II\tablefootmark{11}                    & Y*O   &  & AB         \\
  BD-16 351               & 02:01:35.6 & -16:10:00 & 10.34 & 7.96 & \verb+~+          & 57.1     & -29.7   & \verb+~+      & COL                                       & Ro*   &  & AB         \\
  2MASS J02303239-4342232 & 02:30:32.4 & -43:42:23 & 10.37 & 7.23 & K5V(e)     & 80.5     & -14.9   & \verb+~+      & COL                                       & pr*   &  & AB         \\
  BD-11 648               & 03:21:49.6 & -10:52:17 & 11.63 & 9.26 & G9         & \verb+~+        & \verb+~+       & \verb+~+      & COL                                       & Ro*   &  & \\
  V* V1221 Tau            & 03:28:14.9 & +04:09:47 & 9.7 & 7.44 & G6V+K3v... & 34.4     & -45.0   & \verb+~+      & COL                                       & BY*   &  AB, C & \\
  2MASS J03573723-0416159 & 03:57:37.2 & -04:16:15 & 10.72 & 8.75 & \verb+~+          & \verb+~+        & \verb+~+       & \verb+~+      & COL                                       & Ro*   &  & \\
  HD 26980                & 04:14:22.5 & -38:19:01 & 9.08 & 7.62 & G3V        & 41.78    & 1.40    & 12.42  & COL                                       & Ro*   &  & \\
  HD 27679                & 04:21:10.3 & -24:32:20 & 9.43 & 7.81 & G2V        & 40.0     & -9.1    & \verb+~+      & COL                                       & pr*   &  & \\
  CD-43 1395              & 04:21:48.6 & -43:17:32 & 10.18 & 8.42 & G7V        & 20.4     & 5.6     & \verb+~+      & COL                                       & Ro*   &  & AB         \\
  2MASS J04272050-4420393 & 04:27:20.4 & -44:20:39 & 10.91 & 8.56 & K1V(e)     & 31.3     & 2.2     & \verb+~+      & COL                                       & pr*   &  & AB         \\
  CD-36 1785              & 04:34:50.7 & -35:47:21 & 10.84 & 8.59 & K1Ve       & 34.8     & -1.6    & \verb+~+      & COL                                       & Ro*   &  & \\
  BD+08 742               & 04:42:32.0 & +09:06:00 & 11.19 & 9.12 & G5V:       & 27.60    & -25.50  & \verb+~+      & COL                                       & *     &  & \\
  HD 31242                & 04:51:53.5 & -46:47:13 & 9.85 & 8.16 & G5V        & 32.7     & 14.2    & \verb+~+      & COL                                       & Ro*   &  & SB1        \\
  HD 272836               & 04:53:05.1 & -48:44:38 & 10.72 & 8.24 & K2V(e)     & 30.8     & 17.0    & \verb+~+      & COL                                       & Ro*   &  & \\
  TYC 9178-284-1          & 06:55:25.1 & -68:06:21 & 11.91 & 8.94 & K4Ve       & 2.7      & 20.2    & \verb+~+      & COL                                       & Ro*   &  & \\
  2MASS J08240598-6334024 & 08:24:05.9 & -63:34:02 & 9.87 & 8.13 & G5V        & -15.6    & 22.4    & \verb+~+      & COL                                       & pr*   & AB, C & \\
  V* V479 Car             & 09:23:34.9 & -61:11:35 & 10.16 & 7.96 & K1V(e)     & -28.30   & 17.98   & 11.36  & COL                                       & BY*   &  &         \\
  2MASS J09322609-5237396 & 09:32:26.0 & -52:37:39 & 10.86 & 8.84 & G8V(e)     & -16.5    & 10.9    & \verb+~+      & COL                                       & pr*   &  & \\
  HD 174656               & 18:53:05.9 & -36:10:22 & 9.87 & 7.28 & K0IV(e)    & 0.20     & -21.40  & \verb+~+      & CRA                                       & Ro*   & AB\tablefootmark{2}                            & \\
  V* DZ Cha               & 11:49:31.8 & -78:51:01 & 12.0 & 8.49 & M0Ve       & -38.0    & -8.3    & \verb+~+      & ECH                                       & Or*   &  & \\
  2MASS J11594226-7601260 & 11:59:42.2 & -76:01:26 & 11.31 & 8.3 & K4Ve       & -40.5   & -5.83   & 9.89   & ECH                                       & TT*   &  & \\
  HD 104467               & 12:01:39.1 & -78:59:16 & 8.56 & 6.85 & G3V(e)     & -42.8    & -4.0    & \verb+~+      & ECH                                       & TT*   &  & SB1        \\
  HD 105923               & 12:11:38.1 & -71:10:36 & 9.16 & 7.18 & G8V        & -38.9    & -8.3    & \verb+~+      & ECH                                       & Ro*   &  & AB         \\
  2MASS J12194369-7403572 & 12:19:43.6 & -74:03:57 & 12.5 & 8.86 & M1         & -42.1    & -11.7   & \verb+~+      & ECH                                       & TT*   &  & \\
  2MASS J12202177-7407393 & 12:20:21.7 & -74:07:39 & 12.0 & 8.37 & M1V        & -41.0    & -6.5    & \verb+~+      & ECH                                       & TT*   &  & AB?         \\
  2MASS J12392124-7502391 & 12:39:21.2 & -75:02:39 & 10.3 & 7.78 & K3Ve       & -43.6    & -11.1   & \verb+~+      & ECH                                       & TT*   &  & \\
  TYC 9245-535-1          & 12:56:08.3 & -69:26:53 & 12.06 & 7.99 & K7Ve       & -33.6    & -7.6    & \verb+~+      & ECH                                       & Ro*   &  & AB         \\
  CD-69 1055              & 12:58:25.5 & -70:28:49 & 9.67 & 7.55 & K0Ve       & -43.1    & -18.7   & \verb+~+      & ECH                                       & Ro*   &  & \\
  V* MP Mus               & 13:22:07.5 & -69:38:12 & 9.75 & 7.29 & K1Ve       & -40.80   & -23.30  & \verb+~+      & ECH                                       & TT*   &  & \\
  2MASS J12375231-5200055 & 12:37:52.2 & -52:00:05 & 10.68 & 6.02 & M3V(e)     & -1032.41 & 30.39   & 103.18 & HYA\tablefootmark{12}                      & pr*   &  & \\
  2MASS J22420163+0946091 & 22:42:01.6 & +09:46:09 & 11.99 & 8.26 & K6Ve       & 17.27    & 30.22   & 26.73  & HYA\tablefootmark{12}                      & pr*   &  & \\
  2MASS J23350028+0136193 & 23:35:00.2 & +01:36:19 & 9.59 & 6.04 & K7V        & 341.31   & 25.50   & 52.56  & IC2391\tablefootmark{12}                   & pr*\tablefoottext{c}   & A*F,BC\tablefootmark{1}                        & \\
  TYC 8639-1114-1         & 11:55:42.9 & -56:37:31 & 12.11 & 8.04 & M0Ve       & -45.4    & -6.0    & \verb+~+      & LCC                                       & Ro*   &  & AB         \\
  TYC 8978-3494-1         & 12:12:48.8 & -62:30:31 & 12.13 & 7.96 & K7Ve       & -36.4    & -16.4   & \verb+~+      & LCC                                       & Ro*   &  & AB         \\
  TYC 8637-1558-1         & 12:16:01.1 & -56:14:07 & 11.22 & 7.96 & K5Ve       & -42.7    & -10.3   & \verb+~+      & LCC                                       & Ro*   &  & \\
  CD-62 657               & 12:28:25.3 & -63:20:58 & 9.08 & 7.33 & G7V        & -34.9    & -11.3   & \verb+~+      & LCC                                       & Ro*   &  & \\
  CPD-63 2367             & 12:36:38.9 & -63:44:43 & 9.69 & 7.37 & K2V        & -37.8    & -9.7    & \verb+~+      & LCC                                       & Ro*   & A*P\tablefootmark{1}                           &         \\
  2MASS J12474824-5431308 & 12:47:48.2 & -54:31:30 & 11.82 & 8.05 & M0Ve       & -41.5    & -11.0   & \verb+~+      & LCC                                       & pr*   &  & \\
  HD 112245               & 12:56:09.4 & -61:27:25 & 9.63 & 7.26 & K0Ve       & -52.5    & -14.7   & \verb+~+      & LCC                                       & Ro*   &  & AB         \\
  TYC 9012-1005-1         & 13:44:42.7 & -63:47:49 & 11.04 & 7.74 & K4Ve       & -44.9    & -20.8   & \verb+~+      & LCC                                       & Ro*   & A,Z*r,A*Y\tablefootmark{1}                     & \\
  CPD-64 1859             & 12:19:21.6 & -64:54:10 & 9.87 & 7.4 & K3V        & -40.20   & -10.20  & \verb+~+      & LCC\tablefootmark{15}                      & Ro*   &  & \\
  TYC 7846-833-1          & 15:56:44.0 & -42:42:29 & 11.88 & 8.27 & K6IVe      & -13.7    & -16.6   & \verb+~+      & LUP                                       & Ro*   &  & AB, C         \\
  CD-33 11275             & 16:35:22.4 & -33:28:53 & 11.42 & 8.08 & K4Ve       & -16.8    & -33.8   & \verb+~+      & LUP                                       & Ro*   &  &      AB, C    \\
  CD-58 860               & 04:11:55.6 & -58:01:47 & 10.01 & 8.36 & G6V        & -7.9     & 38.2    & \verb+~+      & OCT                                       & Ro*   &  & \\
  2MASS J04302731-4248466 & 04:30:27.3 & -42:48:46 & 10.75 & 8.73 & G9V(e)     & 2.6      & 20.8    & \verb+~+      & OCT                                       & pr*   &  & AB         \\
  HD 271037               & 05:06:50.5 & -72:21:11 & 10.91 & 8.67 & K0IV       & -7.0     & 24.7    & \verb+~+      & OCT                                       & Ro*   &  & \\
  HD 274576               & 05:28:51.3 & -46:28:19 & 10.5 & 8.81 & G6V        & -10.3    & 22.6    & \verb+~+      & OCT                                       & Ro*   &  & AB         \\
  BD-20 1111              & 05:32:29.3 & -20:43:33 & 10.4 & 8.7 & \verb+~+          & -8.3     & 12.0    & \verb+~+      & OCT                                       & *     &  & Ba, Bb ?        \\
  CD-47 1999              & 05:43:32.1 & -47:41:10 & 10.19 & 8.64 & G0V        & -10.0    & 15.0    & \verb+~+      & OCT                                       & Ro*   &  & AB?         \\
  2MASS J05581182-3500496 & 05:58:11.8 & -35:00:49 & 11.24 & 9.38 & G9V        & -12.7    & 13.0    & \verb+~+      & OCT                                       & pr*   &  & \\
  2MASS J06033540-4911256 & 06:03:35.4 & -49:11:25 & 11.18 & 9.08 & K0Ve       & -9.3     & 8.3     & \verb+~+      & OCT                                       & pr*   &  & AB         \\
  CD-66 395               & 06:25:12.3 & -66:29:10 & 10.87 & 9.0 & K0IV       & -18.40   & 24.40   & \verb+~+      & OCT                                       & Ro*   &  & \\
  2MASS J06400573-3033089 & 06:40:05.7 & -30:33:08 & 10.24 & 8.7 & F9V        & -12.60   & 7.70    & \verb+~+      & OCT                                       & pr*   & AB\tablefootmark{2}                            & \\
  BD-18 4452             & 17:13:11.6 & -18:34:25 & 10.86 & 6.48 & M0Ve       & 39.0     & -88.7   & \verb+~+      & OCT                                       & pr*   & AB\tablefootmark{1} & A, Ba, Bb         \\
CD-30 3394 & 06:40:04.9 & -30:33:03  & 9.83 & 8.59 & F6V & -18.3 & 7.0 & \verb+~+ & OCT & * &  & A, Ba, Bb \\
  V* BO Mic               & 20:47:45.0 & -36:35:40 & 9.39 & 6.79 & K3V(e)     & 11.42    & -75.87  & 19.16  & PMG\tablefootmark{12}                      & BY*   &  & \\
  V* V1002 Sco            & 16:12:40.5 & -18:59:28 & 9.81 & 7.49 & K6e        & -11.70   & -22.20  & \verb+~+      & RHO                                       & Or*   & AB,Aa\tablefootmark{1}                         & AB         \\
  2MASS J16141107-2305362 & 16:14:11.0 & -23:05:36 & 10.5 & 7.46 & K2IV       & -11.9    & -22.9   & \verb+~+      & RHO                                       & TT*   & A*G,Aa,Ab\tablefootmark{1}                     & \\
  HD 147808               & 16:24:51.3 & -22:39:32 & 9.5 & 7.08 & G9IVe      & -13.4    & -20.0   & \verb+~+      & RHO                                       & TT*   & AB,Aa,Ab\tablefootmark{1}                      & \\
  V* V896 Sco             & 16:25:38.4 & -26:13:54 & 11.75 & 7.52 & M          & \verb+~+        & \verb+~+       & \verb+~+      & RHO                                       & Or*   &  & AB         \\
  2MASS J16312019-2430009 & 16:31:20.1 & -24:30:00 & 10.69 & 7.09 & K3e        & -5.1     & -28.8   & \verb+~+      & RHO                                       & SB*   & AB                     & AB, C         \\
  V* V877 Sco             & 16:33:41.9 & -25:23:34 & 10.66 & 7.95 & K4IVe      & -2.4     & -25.4   & \verb+~+      & RHO                                       & bL*   &  & \\
  2MASS J16471358-1514275 & 16:47:13.5 & -15:14:27 & 12.02 & 8.04 & M0Ve       & -5       & -11     & \verb+~+      & RHO\tablefootmark{17}                      & TT*   &  & AB         \\
  2MASS J16482187-1410427 & 16:48:21.8 & -14:10:42 & 14.79 & 7.86 & \verb+~+          & \verb+~+        & \verb+~+       & \verb+~+      & RHO\tablefootmark{17}                      & Y*O   &  & AB         \\
  CD-78 24                & 00:42:20.3 & -77:47:39 & 10.21 & 7.53 & K3Ve       & 79.5     & -32.1   & \verb+~+      & THA                                       & Ro*   &  & \\
  CD-53 544               & 02:41:46.8 & -52:59:52 & 10.22 & 6.76 & K6Ve       & 97.5     & -13.7   & \verb+~+      & THA                                       & Ro*   &  & \\
  CD-35 1167              & 03:19:08.6 & -35:07:00 & 11.0 & 7.72 & K7Ve       & 89.2     & -20.3   & \verb+~+      & THA                                       & Ro*   &  & \\
  CD-46 1064              & 03:30:49.0 & -45:55:57 & 9.79 & 7.1 & K3V        & 87.2     & -5.0    & \verb+~+      & THA                                       & Ro*   &  & \\
  CD-44 1173              & 03:31:55.6 & -43:59:13 & 10.9 & 7.47 & K6Ve       & 90.9     & -5.0    & \verb+~+      & THA                                       & Ro*   &  &  AB        \\
  HD 22705                & 03:36:53.4 & -49:57:28 & 7.65 & 6.14 & G2V        & 89.74    & 0.29    & 23.07  & THA                                       & *i*   & AB,AC\tablefootmark{1}                         & AB         \\
  BD-12 943               & 04:36:47.1 & -12:09:20 & 9.86 & 7.76 & \verb+~+          & 45.9     & -20.4   & \verb+~+      & THA                                       & Ro*   &  & \\
  TYC 8083-455-1          & 04:48:00.6 & -50:41:25 & 11.53 & 7.92 & K7Ve       & 64.3     & 16.6    & \verb+~+      & THA                                       & Ro*   &  & AB         \\
  BD-19 1062              & 04:59:32.0 & -19:17:41 & 10.61 & 8.07 & \verb+~+          & 42.6     & -10.0   & \verb+~+      & THA                                       & Ro*   &  & \\
  ASAS J051536-0930.8     & 05:15:36.4 & -09:30:51 & 9.82 & 8.08 & G5         & 31.4     & -20.4   & \verb+~+      & THA                                       & TT*   &  & \\
  2MASS J05182904-3001321 & 05:18:29.0 & -30:01:32 & 11.66 & 8.3 & K4Ve       & 37.7     & 4.3     & \verb+~+      & THA                                       & pr*   &  & AB         \\
  TYC 8098-414-1          & 05:33:25.5 & -51:17:13 & 11.74 & 8.16 & K7Ve       & 54.1     & 16.6    & \verb+~+      & THA                                       & Ro*   &  & \\
  2MASS J05490656-2733556 & 05:49:06.5 & -27:33:55 & 10.22 & 8.25 & G9V        & 30.90    & -4.70   & \verb+~+      & THA                                       & pr*   &  & \\
  TYC 9344-293-1          & 23:26:10.7 & -73:23:49 & 11.83 & 7.94 & M0Ve       & 71.5     & -66.5   & \verb+~+      & THA                                       & Ro*   &  & \\
  HD 30051                & 04:43:17.2 & -23:37:42 & 7.12 & 6.02 & F2/3IV/V   & 50.25    & -11.84  & 15.73  & THA\tablefootmark{19}                      & *     &  & \\
  CD-46 10045             & 15:25:11.6 & -46:59:12 & 10.08 & 7.78 & G9V        & -49.63   & -21.24  & 15.46  & UCL                                       & **    & AB\tablefootmark{3}                            & AB         \\
  V* KW Lup               & 15:45:47.6 & -30:20:55 & 9.28 & 6.46 & K2V        & -69.11   & -98.40  & 24.78  & UCL\tablefootmark{16}                      & BY*   &  & AB          \\
  2MASS J15594951-3628279 & 15:59:49.5 & -36:28:27 & 10.36 & 8.03 & K3e        & -16.21   & -51.01  & 15.40  & US                                        & pr*   &  & \\
  V* V857 Ara             & 17:18:14.6 & -60:27:27 & 9.48 & 7.53 & G8V        & -54.62   & -91.04  & 16.97  & US, THA\tablefootmark{16}                  & BY*   &  & AB         \\
  TYC 6234-1287-1         & 17:26:56.5 & -16:31:34 & 11.59 & 7.83 & K4Ve       & -10.1    & -39.0   & \verb+~+      & US?\tablefootmark{13}                      & Ro*   &  & AB, SB2?   \\
  2MASS J09551508-6203324 & 09:55:15.0 & -62:03:32 & 12.01 & 9.66 & K2V        & -15.3    & 4.1     & \verb+~+      & N                                      & pr*   &  & \\
  BD+01 3657              & 18:22:17.2 & +01:42:25 & 10.2 & 6.86 & K6Vke      & 84.65    & -18.29  & 37.28  & N                                      & Ro*   &  & \\
  HD 83096                & 09:31:24.8 & -73:44:49 & 7.3 & 6.17 & F1V        & -35.40   & 32.44   & 11.98  & N                                      & **    &  & \\
  HD 154361               & 17:05:08.4 & -01:47:10 & 9.5 & 6.81 & K0         & 75.7     & -44.2   & \verb+~+      & N                                      & cC*   &  & \\
  TYC 438-902-1           & 18:03:17.8 & +04:48:26 & 10.3 & 8.01 & K2         & -11.8    & -27.0   & \verb+~+      & N                                      & Ro*   &  & \\
  TYC 458-290-1           & 18:31:59.1 & +05:40:19 & 10.82 & 8.85 & \verb+~+          & 31.8     & -5.9    & \verb+~+      & N                                      & Ro*\tablefootmark{b}   &  & \\
  TYC 460-624-1           & 18:45:10.2 & +06:20:15 & 10.76 & 6.81 & \verb+~+          & -38.0    & -83.7   & \verb+~+      & N                                      & Ro*   &  & \\
  TYC 1026-1952-1         & 18:47:25.5 & +08:41:07 & 12.1 & 8.97 & \verb+~+          & -11.1    & -18.2   & \verb+~+      & N                                      & Ro*   &  & AB         \\
  BD+01 3828              & 18:57:19.4 & +01:20:33 & 9.56 & 7.79 & G0         & 27.4     & -0.9    & \verb+~+      & N                                      & *i*   & AB\tablefootmark{2}                            & \\
  HD 176898               & 19:02:17.0 & +02:44:21 & 9.19 & 7.16 & G          & 24.6     & -30.2   & \verb+~+      & N                                      & *     &  & \\
  TYC 490-110-1           & 19:31:34.9 & +05:56:07 & 11.13 & 7.73 & K3III      & 22.2     & 2.8     & \verb+~+      & N                                      & Ro*   &  & AB         \\
  TYC 482-1106-1          & 19:36:04.3 & +03:25:48 & 10.43 & 8.33 & \verb+~+          & 3.0      & -24.7   & \verb+~+      & N                                      & Ro*   &  & \\
  TYC 1062-1904-1         & 19:47:39.1 & +09:29:36 & 11.81 & 8.85 & \verb+~+          & 26.3     & 13.2    & \verb+~+      & N                                      & *     &  & \\
  TYC 1058-1925-1         & 19:51:13.3 & +08:42:19 & 10.48 & 8.13 & \verb+~+          & 11.1     & -17.5   & \verb+~+      & N                                      & *     &  & AB         \\
  TYC 5188-193-1          & 20:35:27.1 & -07:06:54 & 11.59 & 10.78 & \verb+~+          & 16.2     & -14.6   & \verb+~+      & N                                      & *     &  & \\
  HD 199602               & 20:58:32.5 & -08:50:22 & 9.19 & 6.4 & K0         & -31.9    & -56.1   & \verb+~+      & N                                      & *     &  & \\
  TYC 538-573-1           & 21:01:24.5 & +05:42:12 & 11.55 & 9.39 & \verb+~+          & 18.7     & -3.3    & \verb+~+      & N                                      & V*    &  & AB         \\
  TYC 553-84-1            & 21:56:22.0 & +05:17:48 & 10.68 & 6.9 & \verb+~+          & -8.2     & -6.3    & \verb+~+      & N                                      & *     &  & \\
  CCDM J00002+0146AB      & 00:00:12.2 & +01:46:17 & 10.78 & 7.93 & G9IV       & 44.10    & -22.90  & \verb+~+      & N                                      & **    & A\tablefootmark{2}                             & AB, SB3    \\
  TYC 1-1187-1            & 00:09:21.7 & +00:38:06 & 11.88 & 8.71 & K4Ve       & 119.4    & -29.7   & \verb+~+      & N                                      & Ro*   &  & SB2   \\
  HD 531                  & 00:09:51.6 & +08:27:11 & 9.35 & 7.57 & G6V+G7V    & 56.21    & -11.41  & 13.11  & N                                      & **    & AB\tablefootmark{3}                            & AB          \\
  GJ 3029                 & 00:21:37.2 & -46:05:33 & 12.22 & 7.45 & M3Ve       & -302.66  & -360.97 & 51.80  & N                                      & Fl*   &  & \\
  BD+07 85                & 00:38:59.2 & +08:28:41 & 10.09 & 7.2 & K4Ve       & 100.9    & -48.9   & \verb+~+      & N                                      & Ro*   &  & \\
  2MASS J01452133-3957204 & 01:45:21.3 & -39:57:20 & 11.39 & 7.59 & M1V        & 279.02   & 131.42  & 33.84  & N                                      & pr*   & AB\tablefootmark{7}                             & \\
  BD-21 332               & 01:53:11.3 & -21:05:43 & 11.73 & 7.14 & M2Ve       & 271.4    & 72.5    & \verb+~+      & N                                      & Ro*   & SB2\tablefootmark{8}                           & \\
  GJ 3148 A               & 02:16:41.1 & -30:59:18 & 12.05 & 7.13 & M3Ve       & 683.07   & 250.88  & 69.81  & N                                      & Fl*   & AB\tablefootmark{3}                            & \\
  2MASS J14110874-6155469 & 14:11:08.7 & -61:55:47 & 10.8 & 6.92 & M1Ve       & 147.9    & -51.0   & \verb+~+      & N                                      & pr*   &  & \\
  2MASS J15011648-4339311 & 15:01:16.5 & -43:39:31 & 10.8 & 6.07 & M3Ve       & 98.0     & 26.9    & \verb+~+      & N                                      & pr*   &  & SB2        \\
  2MASS J15104047-5248189 & 15:10:40.4 & -52:48:19 & 12.26 & 6.85 & M2V(e)     & -145.3   & -167.5  & \verb+~+      & N                                      & pr*   &  & AB, SB3    \\
  2MASS J15240033-7054096 & 15:24:00.3 & -70:54:09 & 11.93 & 7.77 & M1Ve       & -145.1   & -128.2  & \verb+~+      & N                                      & pr*   &  & SB2        \\
  CD-24 12794             & 16:43:56.9 & -25:08:36 & 11.26 & 7.42 & K6Ve       & 8.7      & -50.7   & \verb+~+      & N                                      & Ro*   &  & \\
  TYC 6242-104-1          & 17:21:56.0 & -20:10:51 & 13.22 & 9.15 & K5Ve       & -11.7    & -13.7   & \verb+~+      & N                                      & Ro*   &  & \\
  V* V2384 Oph            & 17:42:30.3 & -28:44:55 & 9.03 & 7.32 & G5V        & -2.83    & -40.87  & 12.82  & N                                      & bL*   &  & \\
  HIP 96515               & 19:37:08.7 & -51:34:00 & 12.29 & 7.96 & M1Ve       & 91.07    & -21.94  & 22.79  & N                                      & Al*   & AB\tablefootmark{4}                            & AB, SB2    \\
  GJ 781.1 A              & 20:07:44.9 & -31:45:14 & 13.08 & 7.4 & M4Ve       & 286.55   & -750.5 & 67.08  & N                                      & Fl*   & AB\tablefootmark{3}                            & SB2        \\
  HIP 106043              & 21:28:44.4 & -47:15:42 & 12.2 & 8.29 & M1Ve       & -16.42   & -58.10  & 27.94  & N                                      & Ro*   &  & \\
  GJ 841 A                & 21:57:41.2 & -51:00:22 & 10.5 & 5.88 & M2Ve       & -35.66   & -377.25 & 62.61  & N                                      & Fl*   & AB\tablefootmark{3}                            & \\
  BD+02 4456              & 21:59:59.9 & +03:02:24 & 10.35 & 7.89 & K0IV       & 63.80    & -3.20   & \verb+~+      & N                                      & Ro*   &  & SB2        \\
  2MASS J22023015-0406117 & 22:02:30.1 & -04:06:11 & 10.06 & 7.97 & K2V        & 29.9     & -13.4   & \verb+~+      & N                                      & pr*   &  & \\
  V* CS Gru               & 22:15:35.2 & -39:00:50 & 9.32 & 7.09 & K0V        & 94.36    & -39.79  & 17.84  & N                                      & BY*   &  & SB1        \\
  HD 212781               & 22:26:53.2 & -00:50:39 & 10.46 & 8.34 & K0V(e)     & -4.30    & -48.00  & \verb+~+      & N                                      & Ro*   &  & AB, SB1    \\
  HIP 113001              & 22:53:05.9 & +07:28:19 & 11.72 & 7.94 & K6Ve       & 59.38    & -79.98  & 19.62  & N                                      & Ro*   &  & SB2        \\
  TYC 572-382-1           & 22:56:49.5 & +02:35:39 & 10.29 & 8.08 & K1IV       & -14.2    & -18.0   & \verb+~+      & N                                      & Ro*   &  & AB         \\
  2MASS J23045301+0632003 & 23:04:53.0 & +06:32:00 & 10.99 & 8.18 & K1Ve       & -14.8    & -23.4   & \verb+~+      & N                                      & pr*   &  & \\
  2MASS J23085046+0000528 & 23:08:50.4 & +00:00:52 & 10.5 & 8.62 & G8V        & 16.8     & -7.1    & \verb+~+      & N                                      & pr*   &  & \\
  TYC 584-343-1           & 23:21:56.3 & +07:21:32 & 10.97 & 8.94 & K0V        & 21.50    & -2.50   & \verb+~+      & N                                      & Ro*   &  & AB         \\
  V* BS Psc               & 23:52:24.3 & -00:55:59 & 10.76 & 8.95 & G7V        & 19.90    & -12.30  & \verb+~+      & N                                      & BY*   &  & \\
  CD-36 14261             & 20:36:08.3 & -36:07:11 & 11.73 & 7.17 & M3Ve       & 11.4     & 40.3    & \verb+~+      & N                     & Ro*   &  & \\
\hline\\
\end{longtable}
\tablebib{(1)~\cite{Mason2001}; (2)~\cite{Dommanget2002}; (3)~\cite{Turon1993}; (4)~\cite{Dommanget2000}; (5)~\cite{Anderson2012}; (6)~\cite{Gliese1991}; (7)~\cite{Bergfors2010}; (8)~\cite{Shkolnik2010}; (9)~\cite{Kraus2007}; (10)~\cite{Messina2010}; (11)~\cite{Alcala2008}; (12)~\cite{Montes2001}; (13)~\cite{Aarnio2008}; (14)~\cite{Riedel2014}; (15)~\cite{Mamajek2006}; (16)~\cite{Song2012}; (17)~\cite{Sartori2003}; (18)~\cite{Lopez2006}; (19)~\cite{Zuckerman2011}; (20)~\cite{Zuckerman2004}; (21)~\cite{Schlieder2012}}
\tablefoot{\tablefoottext{a}{Formats as 'A*Y' signifies  multiple components AB, AC, ..., AY}
\tablefoottext{b}{Note in SIMBAD to potential component within 30$\arcsec$}
\tablefoottext{c}{Note in SIMBAD for triple system identified by \cite{Malogolovets2007}}
\tablefoottext{d}{SB systems denote spectroscopic binary systems identified in \cite{Torres2008, Elliott2014}}.}
\end{landscape}
}

{
\onecolumn
\section{Individual detections in this work}
\label{sec:appendix_b}
\tiny
\LTcapwidth=\textwidth
\begin{longtable}{p{3.5cm}}
\caption{Details of individual detections in this work.} \label{tab:all_individ_detections}\\
\endfirsthead
\end{longtable}
}
\twocolumn

\onecolumn
\section{Non SACY targets}
\label{sec:non_sacy_targets}
{\tiny
\begin{table}
\center
\caption{A summary of the properties of non-SACY multiple systems identified in this work.}
\label{tab:binaries_2}
\begin{tabular}{p{4cm} p{1.3cm} l l l l l p{1.4cm} p{1.4cm} l l}
\hline\hline\\[-1ex]
  \multicolumn{1}{l}{ID} &
  \multicolumn{1}{l}{Ang. sep.} &
  \multicolumn{1}{l}{PA} &  
  \multicolumn{1}{l}{Dist.} &
  \multicolumn{1}{l}{Phys. sep} &
  \multicolumn{1}{l}{$\Delta$K} &
  \multicolumn{1}{l}{Comp.} &
  \multicolumn{1}{l}{Prob.} &
  \multicolumn{1}{l}{Pop.} &
  \multicolumn{1}{l}{q} &
  \multicolumn{1}{l}{$\sigma_\mathrm{q}$} \\
  \multicolumn{1}{l}{} &
  \multicolumn{1}{l}{($\arcsec$)} &
  \multicolumn{1}{l}{(deg.)} &    
  \multicolumn{1}{l}{(pc)} &
  \multicolumn{1}{l}{(au)} &
  \multicolumn{1}{l}{(mag.)} &
  \multicolumn{1}{l}{} &
  \multicolumn{1}{l}{} &
  \multicolumn{1}{l}{} &
  \multicolumn{1}{l}{} &
  \multicolumn{1}{l}{} 
  \\[-0.1ex]
  \hline\\[-1ex]
  \multicolumn{10}{c}{PMS populations} \\
\hline\\[-1ex]
CPD-53 1875              &     0.21  &  288.6  &  -      &  -         &    2.18  &  A, B  &  0.91  &  CAR\tablefootmark{1}    &  0.27  &  0.08   \\
2MASS J12571172-7640111  &     0.28  &  150.1  &  -      &  -         &    0.69  &  A, B  &  -     &  CHAII\tablefootmark{2}  &  0.61  &  0.05   \\
HD 112245                &     2.56  &  169.4  &  66.0   &  168.9     &    2.48  &  A, B  &  -     &  LCA, C                  &  0.22  &  0.12   \\
TYC 8978-3494-1          &     0.43  &   41.2  &  72.5   &  31.2      &    0.13  &  A, B  &  -     &  LCA, C                  &  0.93  &  0.2   \\
TYC 8639-1114-1          &     0.84  &  305.6  &  79.0   &  66.4      &    0.42  &  A, B  &  -     &  LCA, C                  &  0.80  &  0.2   \\
CD-33 11275              &     0.18  &  252.9  &  123.7  &  22.3      &    0.04  &  A, B  &  -     &  LUP                     &  1.03  &  0.2   \\
CD-33 11275              &     1.78  &    3.7  &  123.7  &  220.2     &    0.54  &  A, C  &  -     &  LUP                     &  0.67  &  0.13   \\
TYC 7846-833-1           &     0.2   &  192.5  &  139.7  &  27.9      &    0.76  &  A, B  &  -     &  LUP                     &  0.58  &  0.11   \\
TYC 7846-833-1           &     6.0   &   19.8  &  139.7  &  837.9     &    2.47  &  A, C  &  -     &  LUP                     &  0.17  &  0.04   \\
V* V896 Sco              &     0.87  &   20.9  &  125.0  &  108.7     &    1.68  &  A, B  &  -     &  RHO                     &  0.30  &  0.05   \\
V* V1002 Sco             &     0.12  &  156.0  &  138.7  &  16.6      &    1.58  &  A, B  &  0.89  &  RHO                     &  0.32  &  0.03   \\
2MASS J16312019-2430009  &     0.32  &  170.4  &  103.4  &  33.1      &    2.98  &  A, B  &  -     &  RHO                     &  0.12  &  0.05   \\
2MASS J16312019-2430009  &     4.48  &   12.1  &  103.4  &  463.2     &    0.33  &  A, C  &  -     &  RHO                     &  0.79  &  0.14   \\
2MASS J16471358-1514275  &     0.35  &   60.5  &  -      &  -         &    1.15  &  A, B  &  -     &  RHO\tablefootmark{3}    &  0.44  &  0.05   \\
2MASS J16482187-1410427  &     1.01  &  252.2  &  -      &  -         &    1.93  &  A, B  &  -     &  RHO\tablefootmark{3}    &  0.26  &  0.05   \\
CD-46 10045              &     0.62  &  334.5  &  67.9   &  42.1      &    0.07  &  A, B  &  -     &  UCL                     &  0.96  &  0.2   \\
V* V857 Ara              &     0.2   &  194.2  &  -      &  -         &    2.45  &  A, B  &  -     &  US                      &  0.19  &  0.06   \\
TYC 6234-1287-1          &     1.51  &  245.2  &  -      &  -         &    0.21  &  A, B  &  -     &  US?\tablefootmark{4}    &  0.88  &  0.05   \\ 

\hline\\[-1ex]
  \multicolumn{10}{c}{Other} \\
\hline\\[-1ex]
 HIP 96515                &     8.58  &  224.0  &  -      &  -         &    6.36  &  A, B  &  -     &  N, Field                &  -     &  -      \\
HD 212781                &     5.34  &   77.3  &  -      &  -         &    4.71  &  A, B  &  -     &  N, Field                &  -     &  -      \\
TYC 1026-1952-1          &     6.99  &   58.6  &  -      &  -         &    3.76  &  A, B  &  -     &  N, Field                &  -     &  -      \\
TYC 572-382-1            &     3.07  &   53.5  &  -      &  -         &    2.68  &  A, B  &  -     &  N, Field                &  -     &  -      \\
TYC 490-110-1            &     2.42  &  180.3  &  -      &  -         &    2.45  &  A, B  &  -     &  N, Field                &  -     &  -      \\
TYC 538-573-1            &     6.19  &   25.3  &  -      &  -         &    2.27  &  A, B  &  -     &  N, Field                &  -     &  -      \\
TYC 584-343-1            &     5.08  &  228.5  &  -      &  -         &    1.53  &  A, B  &  -     &  N, Field                &  -     &  -      \\
2MASS J15104047-5248189  &     0.47  &  294.7  &  -      &  -         &    1.14  &  A, B  &  -     &  N, Field                &  -     &  -      \\
CCDM J09314-7345AB       &     0.95  &  133.1  &  -      &  -         &    0.62  &  A, B  &  0.99  &  N, Field                &  -     &  -      \\
TYC 1058-1925-1          &     0.76  &  236.1  &  -      &  -         &    0.17  &  A, B  &  -     &  N, Field                &  -     &  -      \\
CCDM J00002+0146AB       &     1.81  &  262.9  &  -      &  -         &    0.16  &  A, B  &  -     &  N, Field                &  -     &  -      \\
HD 531                   &     5.21  &  275.9  &  -      &  -         &    0.05  &  A, B  &  -     &  N, Field                &  -     &  -      \\

\hline\end{tabular}
\tablebib{(1)~\cite{Messina2010}; (2)~\cite{Alcala2008}; (3)~\cite{Sartori2003}; (4)~\cite{Aarnio2008}}
\end{table}
}


\end{appendix}
\end{document}